\documentclass[a4paper,11pt]{article}
\pdfoutput=1

\usepackage[utf8]{inputenc}
\usepackage{a4wide}
\usepackage{graphicx}
\graphicspath{{Figures/}} 
\usepackage[small,bf]{caption}
\usepackage{amssymb}
\usepackage{amsmath}
\usepackage{slashed}
\usepackage{mathtools}
\usepackage{cite} 
\usepackage{hyperref}
\usepackage{xcolor}

\DeclareMathAlphabet{\mathpzc}{OT1}{pzc}{m}{it}

\newcommand{\be}{\begin{equation}}
\newcommand{\ee}{\end{equation}}
\newcommand{\ben}{\begin{eqnarray}}
\newcommand{\een}{\end{eqnarray}}
\newcommand{\bi}{\begin{itemize}}
\newcommand{\ei}{\end{itemize}}

\numberwithin{equation}{section}
\makeatletter
\g@addto@macro\bfseries{\boldmath}
\makeatother

\begin{document}

\begin{titlepage}
\begin{flushright}
IFT-UAM/CSIC-20-111\\ FTUAM-20-14\\
FTUV-20-0717.1122\\ IFIC/20-37 
\end{flushright}
\vspace*{0.8cm}

\begin{center}
{\Large\bf $\nu$ Electroweak Baryogenesis}\\[0.8cm]
E.~Fern\'andez-Mart\'inez,$^a$
J.~L\'opez-Pav\'on,$^b$
T.~Ota$^a$
and S.~Rosauro-Alcaraz$^a$\\[0.4cm]
$^{a}$\,{\it Departamento de F\'isica Te\'orica and Instituto de F\'{\i}sica Te\'orica, IFT-UAM/CSIC,\\
Universidad Aut\'onoma de Madrid, Cantoblanco, 28049, Madrid, Spain} \\
$^{b}$\,{\it Instituto de F\'isica Corpuscular, Universidad de Valencia and CSIC, \\
Edificio Insitutos Investigaci\'on, Catedr\'atico Jos\'e Beltr\'an 2, 46980, Paterna, Spain}\\
\end{center}
\vspace{0.8cm}

\begin{abstract}
We investigate if the CP violation necessary for successful electroweak baryogenesis may be sourced by the neutrino Yukawa couplings. In particular, we consider an electroweak scale Seesaw realization with sizable Yukawas where the new neutrino singlets form (pseudo)-Dirac pairs, as in the linear or inverse Seesaw variants. We find that the baryon asymmetry obtained strongly depends on how the neutrino masses vary within the bubble walls. Moreover, we also find that flavour effects critically impact the final asymmetry obtained and that, taking them into account, the observed value may be obtained in some regions of the parameter space. This source of CP violation naturally avoids the strong constraints from electric dipole moments and links the origin of the baryon asymmetry of the Universe with the mechanism underlying neutrino masses. Interestingly, the mixing of the active and heavy neutrinos needs to be sizable and could be probed at the LHC or future collider experiments.
\end{abstract}
\end{titlepage}

\section{Introduction}
The origin of the baryon asymmetry of the Universe (BAU) remains one of the most intriguing open questions of the Standard Model (SM). It has been measured with great precision through cosmological observations of the cosmic microwave background (CMB)~\cite{Aghanim:2018eyx} and Big Bang nucleosynthesis (BBN) to be
\begin{equation}
     Y_B^{obs}\equiv \frac{n_b-n_{\bar{b}}}{s}\equiv \frac{n_B}{s}\simeq \left(8.59\pm 0.08\right)\times10^{-11},
    \label{Eq:YB_exp}
\end{equation}
where $n_b$ ($n_{\bar{b}}$) is the (anti)baryon number density and $s$ is the entropy density. To dynamically generate the BAU, the three Sakharov conditions~\cite{Sakharov:1967dj} need to be satisfied: baryon number violation, C and CP violation, and departure from thermal equilibrium. The SM in principle satisfies all three conditions; baryon number is violated in the Early Universe through sphaleron effects~\cite{Klinkhamer:1984di, Kuzmin:1985mm}, C is broken by weak interactions, there is CP violation in the Cabibbo-Kobayashi-Maskawa (CKM) quark mixing matrix, and the departure from thermal equilibrium may occur during the electroweak (EW) phase transition. However, the amount of CP violation in the quark sector is not enough to generate the BAU~\cite{Gavela:1993ts, Gavela:1994ds, Gavela:1994dt} and the phase transition is rather a crossover~\cite{Kajantie:1996mn,Degrassi:2012ry} given the observed mass of the Higgs boson.

Thus, a dynamical generation of the BAU requires physics beyond the SM (BSM). As a minimal option, an extension of the SM scalar sector could make Electroweak Baryogenesis (EWBG)~\cite{Shaposhnikov:1987tw, Cohen:1987vi, Cohen:1990it, Cohen:1990py, Nelson:1991ab} viable. In particular, new scalars could induce a strong first order phase transition~\cite{Dine:1990fj, Dine:1991ck, Anderson:1991zb, Espinosa:1993bs, Espinosa:2011ax, Tofighi:2015fia} at the EW scale and also contribute with new sources of CP violation. In this case, all the interesting physics would be around $\mathcal{O}(100)$~GeV, at the reach of the Large Hadron Collider (LHC)~\cite{Profumo:2007wc,No:2013wsa,Profumo:2014opa,Kozaczuk:2014kva,Dorsch:2017nza,Huang:2017jws}. However, new sources of CP violation induce electric dipole moments, which are very tightly  constrained~\cite{ACME_CP}. Thus, EWBG models usually rely on some dark sector to avoid them (see for example Refs.~\cite{Hall:2019ank, Carena:2018cjh, Guo:2016ixx}). 

Given that the experimental evidence for neutrino masses from the observation of the neutrino oscillation phenomenon~\cite{Fukuda:1998mi,Ahmad:2001an,Eguchi:2002dm,Abe:2011fz,An:2012eh,Ahn:2012nd} is also at odds with the SM, it represents another main window to new physics. 
It is therefore interesting to consider whether new sources of CP violation from the neutrino mass mechanism can generate the observed BAU. 
In the context of the type-I Seesaw mechanism~\cite{Minkowski:1977sc,Mohapatra:1979ia,Yanagida:1979as,GellMann:1980vs}, with extra Majorana right-handed (RH) neutrinos close to the Grand Unification scale, this possibility arises naturally through the baryogenesis via leptogenesis framework~\cite{Fukugita:1986hr}. Moreover, the smallness of neutrino masses is very naturally accommodated by the suppression due to the heavy Majorana masses. Unfortunately, the new physics scale is too high to be testable except in a few examples~\cite{Pascoli:2006ci, Moffat:2018smo, Cirigliano:2006nu, Merlo:2018rin}, and the inclusion of such a large scale destabilizes the Higgs mass and worsens the electroweak hierarchy problem~\cite{Vissani:1997ys,Casas:2004gh}.

Low-energy variants of the Seesaw, such as the inverse or the linear Seesaw models~\cite{Mohapatra:1986aw,Mohapatra:1986bd,Bernabeu:1987gr,Malinsky:2005bi}, include instead the new states around the EW scale. In these scenarios, neutrino masses are naturally protected by an approximate lepton number symmetry~\cite{Branco:1988ex,Kersten:2007vk, Abada:2007ux}. However, a sufficiently large baryon asymmetry cannot be generated through the standard leptogenesis mechanism because lepton number violation is too small.\footnote{The leptogenesis scenario in the inverse/linear Seesaw framework has been discussed in Refs.~\cite{Gu:2019nhb,Gu:2019ott}.} Instead, leptogenesis via oscillations would be possible~\cite{Akhmedov:1998qx,Asaka:2005pn}. This mechanism requires the new states to be out of equilibrium and thus typically rather small Yukawa couplings. Nevertheless, future experiments are sensitive to a considerable fraction of the parameter space of these models corresponding to the strong wash out regime~\cite{Akhmedov:1998qx,Asaka:2005pn,Hernandez:2016kel,Chun:2017spz,Abada:2018oly}. In particular, future measurements from neutrino oscillations, beam-dump experiments and neutrinoless double beta decay could in principle provide sufficient information to predict the BAU generated in the minimal $\mathcal{O}(\rm{GeV})$ scale model~\cite{Hernandez:2016kel}.

The lepton number protection present in the inverse or linear variants of the Seesaw mechanism does allow for larger Yukawa couplings and consequently larger mixing between the new heavy states and the active neutrinos, leading to more interesting phenomenology. Therefore, they also naturally possess all the ingredients for EWBG to work: the large neutrino Yukawa couplings can be a source of the CP violation and an extra singlet scalar can generate the heavy neutrino masses around the EW scale and induce the first order phase transition, while avoiding bounds from electric dipole moments. Variants of this idea, but in the context of a type-I Seesaw without the approximate lepton number symmetry, were studied in Refs.~\cite{Cohen:1990it, Long:2017rdo,Hall:2005aq}.

Here we will investigate the viability of EWBG in the context of low-scale Seesaw mechanisms in which neutrino masses are generated from a soft breaking of lepton number. The heavy neutrinos will thus be arranged in (pseudo-)Dirac pairs. In particular, we will explore the possibility to have all the new physics at the EW scale. A new scalar singlet, which can be responsible for the required strong first order phase transition, will also induce the Dirac mass of the heavy neutrinos. 
With these ingredients, a CP asymmetry in the SM neutrinos may be produced through reflections and transmissions with the bubble wall. The imbalance between neutrinos and antineutrinos will then be converted into a baryon asymmetry through sphaleron processes in the unbroken phase. The generated net baryon number then enters the broken phase as the bubbles expand, where sphalerons are no longer efficient and baryon number is frozen out. 
This scenario was originally proposed in Ref.~\cite{Hernandez:1996bu}. We will revisit its results and reconsider some of the assumptions made in Ref.~\cite{Hernandez:1996bu}. In particular, we will study the impact of different wall profiles in the final BAU and also investigate the inclusion of wash-out and flavour effects.  

This paper is organized as follows. In Section~\ref{Sec:Lagrangian} we introduce the lagrangian of the model and identify the new sources of CP violation. In Sections~\ref{Sec:CP_asymmetry} and \ref{Sec:Diff_eq} we study the generation of a CP asymmetry in neutrinos at the bubble wall and discuss its subsequent diffusion and conversion to a baryon asymmetry. Our numerical results are presented in Section~\ref{Sec:Results} and we summarize our findings in Section \ref{Sec:Conclusions}. 

\section{Neutrino mass generation and CP violation}\label{Sec:Lagrangian}
In this section we specify the particle content of the model and the parametrization we will adopt. We also discuss the source of CP violation arising from the neutrino mass generation mechanism. 
The SM is simply extended by three singlet Dirac neutrinos and a scalar singlet:
\begin{equation}
    \mathcal{L}=-\bar{L}_L \tilde{H} Y_{\nu} N_R -\bar{N}_L \phi Y_N N_R  + h.c. - V\left(\phi^* \phi,H^{\dagger}H\right),
    \label{Eq:Lagrangian}
\end{equation}
where $\phi$ is the singlet scalar and $H$ is the Higgs doublet, $L_L$ is the lepton doublet and $N_{R(L)}$ is the right(left)-handed component of the new Dirac neutrinos. $Y_{\nu}$ and $Y_N$ are $3\times3$ Yukawa matrices. 
The manifest lepton number symmetry of the Lagrangian will then be broken by either (or both) a Majorana mass for $N_L$ (as in the inverse Seesaw scenarios) or a Yukawa coupling between $L_L$ and $N_L^c$ (as in the linear Seesaw realizations) so as to generate the small neutrino masses. Notice that the symmetry can also be broken by a Majorana mas term for $N_R$, however the contribution to light neutrino masses arises at the one loop level~\cite{Pilaftsis:1991ug,Grimus:2002nk,LopezPavon:2012zg}. 
We will remain agnostic as to the specifics of the lepton-number-violating contribution responsible for the observed neutrino masses, and in what follows, we will neglect these small perturbations on the underlying lepton-number-conserving structure. 
We only remark that the $3 \times 3$ Majorana mass matrix for $N_L$ and the $3 \times 3$ Yukawa coupling between $L_L$ and $N_L^c$ contain enough degrees of freedom so as to reproduce the correct pattern of neutrino masses and mixings regardless of the values of $Y_\nu$ or $Y_N$. Thus, no conditions on $Y_\nu$ or $Y_N$ can be derived from neutrino oscillation data. 
The last term in Eq.~(\ref{Eq:Lagrangian}) refers to the scalar potential, which couples the Higgs doublet to the singlet scalar and can induce the strong first order phase transition~\cite{Espinosa:1993bs, Espinosa:2011ax, Tofighi:2015fia}. 

After spontaneous symmetry breaking (SSB), we will assume that both the SM Higgs field and the singlet scalar develop a vacuum expectation value (vev), $v_H$ and $v_{\phi}$, respectively, which generate the following Dirac mass terms for the neutrino states:
\begin{equation}
    \mathcal{L}_{mass}=-\bar{\nu}_L m_D N_R - \bar{N}_L M_N N_R  + h.c.,
    \label{Eq:Lag_Mass}
\end{equation}
where we have defined $m_D\equiv v_H Y_{\nu}/\sqrt{2}$ and $M_N\equiv v_{\phi} Y_N$.
As discussed above, lepton number conservation ensures that the three heavy neutrinos $N_i$ have Dirac masses while
the three light neutrinos $\nu_i$ remain massless even for large values of $m_D$ and a low $M_N$ scale\footnote{Their light masses will instead be tied to the small lepton number breaking parameters of the inverse or linear seesaw that can be safely neglected for the generation of baryon asymmetry.}.
The mixing between the heavy neutrinos and the active states is given by the ratio between the Dirac masses 
\begin{equation}
    \theta\equiv m_D M_N^{-1},
    \label{Eq:Mixing}
\end{equation}
and can thus be sizable.
There is only one source of CP violation not suppressed by the generally smaller charged lepton Yukawas~\cite{Hernandez:1996bu,Santamaria:1993ah},
which is associated to the following basis invariant~\cite{Jarlskog:1985ht, Jenkins:2007ip, Jenkins:2009dy}
\begin{equation}
    \delta_{CP}\equiv \text{Im}\text{Tr}\left[M_N^{\dagger} M_N m_D^{\dagger} m_D M_N^{\dagger} M_N M_N^{\dagger} M_N m_D^{\dagger} m_D m_D^{\dagger} m_D\right].
    \label{Eq:Reph_Inv}
\end{equation}
In the basis where $M_N$ is real and diagonal with eigenvalues $M_i$, one finds~\cite{Hernandez:1996bu}
\begin{equation}
    \delta_{CP}=M_1^2 M_2^2 M_3^2(M_1^2-M_2^2)(M_2^2-M_3^2)(M_3^2-M_1^2)\text{Im}\left[(\theta^{\dagger}\theta)_{12}(\theta^{\dagger}\theta)_{23}(\theta^{\dagger}\theta)_{31}\right].
    \label{Eq:Jarlskog_Inv}
\end{equation}

Notice that the CP invariant is suppressed by the sixth power of $\theta$, hence the importance of ensuring large mixing and the reason to consider low scale seesaw realizations in this context that decouple its size from the smallness of neutrino masses. Nevertheless, constraints from precision electroweak and flavour observables exist~\cite{Shrock:1980ct,Shrock:1981wq,Langacker:1988ur,Tommasini:1995ii,Antusch:2006vwa,Antusch:2008tz,Antusch:2014woa,Fernandez-Martinez:2016lgt,Coutinho:2019aiy} on the combination $\theta \theta^\dagger$. Indeed, $\theta \theta^\dagger$ represents the coefficient of the only dimension 6 operator obtained at tree level\footnote{And therefore the least suppressed effective operator in the absence of the Weinberg dimension 5 operator~\cite{Weinberg:1979sa}, which is only induced by the smaller lepton number-violating parameters.} when integrating out the heavy neutrino degrees of freedom~\cite{Broncano:2002rw}. The $d=6$ operator physically leads to deviations from unitarity of the PMNS mixing matrix given the non-negligible mixing with the heavy states. 
These constraints will represent the main limiting factor to the final baryon asymmetry that we will compute in the next sections.

In general, the Dirac mass matrix $m_{D}$ can be parametrized through a bi-unitary transformation as
\begin{equation}
    m_D\equiv U_l m_d V_R^{\dagger},
    \label{Eq:FullParam_mD}
\end{equation}
where $m_d$ is a diagonal matrix with positive real entries $m_{d_\alpha}$ and $U_l$ and $V_R$ are $3\times3$ unitary matrices. 

The physical degrees of freedom of the matrix $V_R$ can be parametrized by three mixing angles and one CP violating phase in complete analogy to the CKM matrix. 
With this parametrization:
\begin{equation}
    \delta_{CP}=(m_{d_e}^2-m_{d_\mu}^2)(m_{d_\mu}^2-m_{d_\tau}^2)(m_{d_\tau}^2-m_{d_e}^2)(M_1^2-M_2^2)(M_2^2-M_3^2)(M_3^2-M_1^2) J,
    \label{Eq:Jarlskog_Inv2}
\end{equation}
where
\begin{equation}
J= \text{Im}(V_{Ri \alpha} V_{Ri \beta}^* V_{Rj \alpha}^* V_{Rj \beta})
    \label{Eq:JarlskogJ}
\end{equation}
is the usual Jarlskog rephasing invariant with $\alpha \neq \beta$ and $i \neq j$. In order to estimate the maximum size of the baryon number asymmetry achievable, in the following sections we will set $J=1$. This choice fixes the matrix $V_R$.

From Eq.~(\ref{Eq:Jarlskog_Inv2}), a significant hierarchy in the values of  $m_{d_\alpha}$ is also desirable so as to maximize $\delta_{CP}$. An advantageous choice is to set one of the three $m_{d_\alpha}$ to zero while the other two differ by a factor $\sqrt{2}$. Their maximum allowed size will be determined by the existing bounds on the product 
\begin{equation}
\theta \theta^\dagger = U_l m_d V_R^\dagger M_N^{-2} V_R m_d U_l^\dagger
\label{eq:d6op}
\end{equation}
mentioned above. These constraints are significantly flavour-dependent (see e.g. Ref~\cite{Fernandez-Martinez:2016lgt}) and, from Eq.~(\ref{eq:d6op}), the flavour structure is controlled by the degrees of freedom in $U_l$. However, the charged lepton Yukawas imply a stronger suppression as a source of CP violation for baryogenesis compared to that of the neutrinos, so we will neglect them in the rest of this work. Hence, the transformation $U_l$, which is part of the PMNS lepton mixing matrix, becomes unphysical in this limit and can be absorbed in a field redefinition. Thus, the most meaningful constraint that can be derived from $\theta \theta^\dagger$ on $m_{d_\alpha}$ is through $Tr[\theta\theta^\dagger]\leq 0.007$~\cite{Fernandez-Martinez:2016lgt} at $2 \sigma$, since this quantity does not depend on $U_l$.

For the sake of definiteness, we will set $U_l=I$ and choose  $m_{d_e} = m_{d_\tau}/\sqrt{2}$ and $m_{d_\mu} = 0$. This choice for $m_{d_\alpha}$ implies that the neutrino Yukawa couplings to the second and third heavy states have the same magnitude. Additionally, it makes the coupling to the muon to vanish, for which the bounds on $\theta \theta^\dagger$ are the most stringent~\cite{Antusch:2014woa,Fernandez-Martinez:2016lgt,Coutinho:2019aiy}. 
Therefore, the matrix $m_D$ now depends on a single parameter, $m_{d_\tau}$.
Note, however, that rotations of this particular choice with other values of $U_l$ would be completely equivalent. In other words, all the ``flavour'' indices in this work will not necessarily correspond to the electron, muon or tau flavours, since  their masses have been neglected. 

In the following sections we will present results as a function of the remaining free parameters of the model. Namely, the three diagonal entries of $M_N$ ($M_i$) which correspond to leading order with the physical masses of the three heavy Dirac neutrinos, as well as $m_{d_\tau}$, respecting the constraints on $Tr[\theta\theta^\dagger] $ through Eq.~(\ref{eq:d6op}) for the different values of $M_i$ considered. 

\section{Generation of a CP asymmetry}\label{Sec:CP_asymmetry}
\begin{figure}
    \centering
    \includegraphics[width=0.80\textwidth]{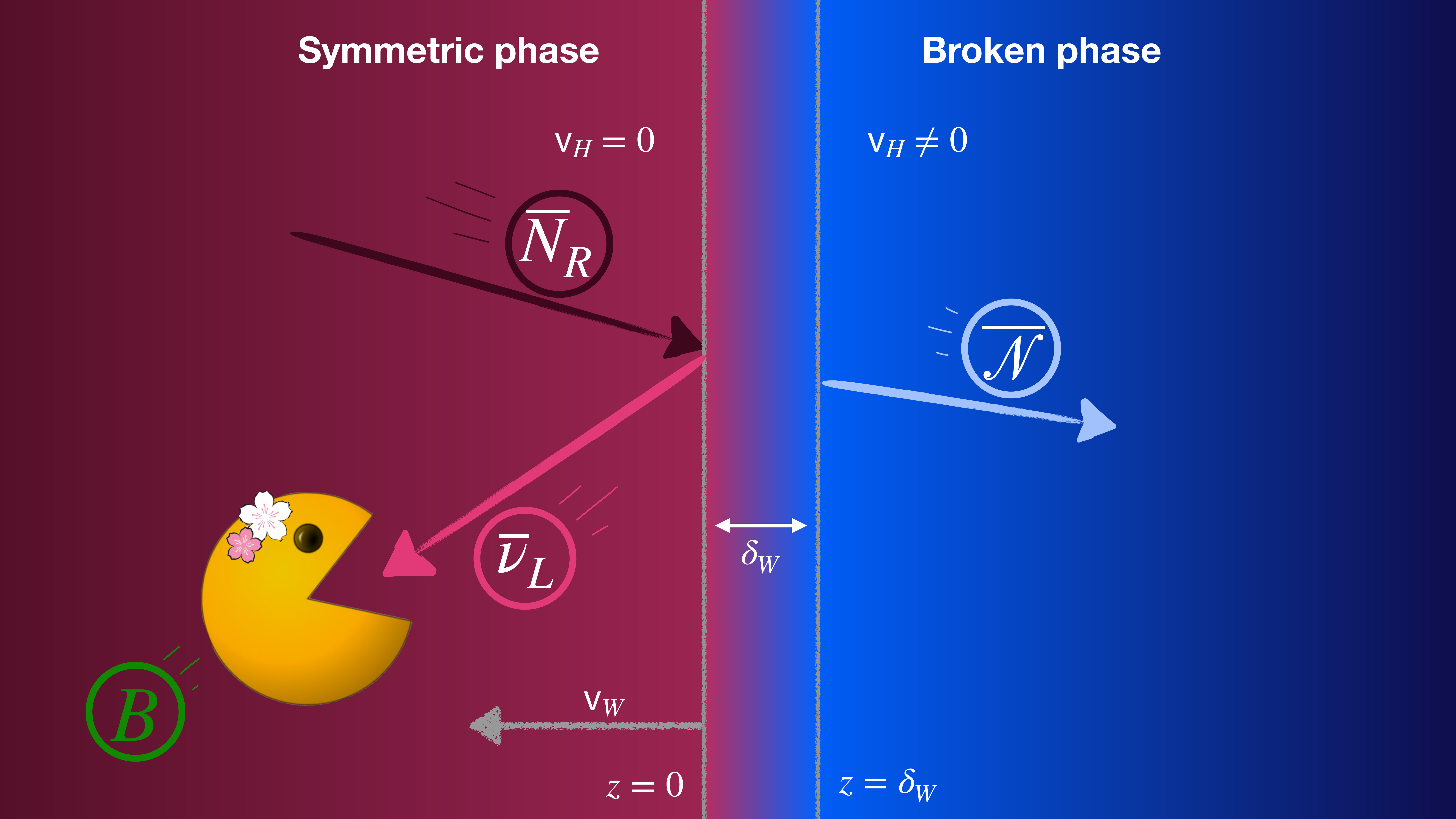}
    \caption{Sketch of the reflection of $\bar{N}_R$ off a bubble whose wall width is $\delta_W$ to $\bar{\nu}_L$ and its subsequent conversion to baryons through shpaleron processes. If there is CP violation, the reflected $\nu_L$ will be different to the $\bar{\nu}_L$ and thus a baryon asymmetry can be generated.}
    \label{Fig:Picture_ref}
\end{figure}

In the presence of the new scalar singlet, a strong first order phase transition is possible~\cite{Espinosa:1993bs, Espinosa:2011ax, Tofighi:2015fia}. Depending on the parameters of the scalar sector and its couplings to fermions, bubbles of a given width $\delta_W$ will start nucleating at the temperature $T_c$ and expand at a velocity $v_W$. 
Neutrinos travelling from the unbroken phase towards the bubble wall will be reflected by the wall as depicted in Fig.~\ref{Fig:Picture_ref}.
In the presence of CP violation, the reflection rate for neutrinos and antineutrinos will be different, generating an asymmetry in $\nu_L$, which will subsequently be converted to a baryon asymmetry through sphaleron transitions.
In the following we will assume that the phase transition is sufficiently strong. Consequently, the sphaleron rate will be suppressed inside the bubbles, such that any baryon asymmetry generated in the symmetric phase will be preserved after entering the regions of true vacuum. 

We will devote the rest of this section to describe the generation of the CP asymmetry through reflections and transmissions of neutrinos in the bubble wall. We assume that the bubbles are sufficiently large such that their surface can be described as a plane and gravitational effects can be neglected~\cite{Coleman:1980aw}.
We closely follow the method developed in Refs.~\cite{Nelson:1991ab, Cline:2000nw} to solve the Dirac equation for the different neutrino species. An asymmetry may be induced through the dependence of their mass matrix on the direction perpendicular to the wall, $z$. 
The $z$-dependence of the mass matrix arises solely from the change in the value of the scalar vevs from the unbroken to the broken phase. Performing a boost to the wall rest frame, only the dependence in the spatial dimension $z$ is relevant. The formal solution to the Dirac equation can be written as~\cite{Cline:2000nw}
\begin{equation}
    \mathcal{N}=e^{-iEt}
    \begin{pmatrix}
    L(z)\\
    R(z)
    \end{pmatrix}
    \otimes \chi_s,
    \label{Eq:Formal_sol}
\end{equation}
where $E$ is the energy of the state and we have separated the chirality states ($L\equiv \begin{pmatrix} \nu_L & N_L\end{pmatrix}^T$ and $R\equiv N_R$) and the spin state $\chi_s$. Spin is conserved upon reflection or transmission, such that $\sigma_3 \chi_s = s\chi_s$. Using this ansatz we find that the chirality states satisfy
\begin{equation}
    \left(is\partial_z + \mathcal{Q}(z)\right)
    \begin{pmatrix}
    L(z)\\
    R(z)
    \end{pmatrix}=0,\hspace{10mm} \mathcal{Q}(z)\equiv
    \begin{pmatrix}
    E & -\mathcal{M}(z)\\
    \mathcal{M}(z)^{\dagger} & -E
    \end{pmatrix},
    \label{Eq:Final_Eq}
\end{equation}
where $\mathcal{M}(z)\equiv\begin{pmatrix} m_D(z) & M_N(z)\end{pmatrix}^T$ and we generally denote with $E$ diagonal submatrices of the appropriate dimension with the energy of the corresponding states. The formal solution
to Eq.~(\ref{Eq:Final_Eq}) at a position $z$ can be expressed as
\begin{equation}
    \begin{pmatrix}
    L(z)\\
    R(z)
    \end{pmatrix}
    = \mathcal{P}\left(e^{\frac{i}{s}\int_{0}^z\mathcal{Q}(z')dz'}\right)
    \begin{pmatrix}
    L(0)\\
    R(0)
    \end{pmatrix},
\end{equation}
where $\mathcal{P}$ denotes the $z$-ordered product and $z=0$ is the position where the bubble wall starts in the unbroken phase. 
Given that the mass matrix only varies within the bubble wall, but is constant inside or outside the bubble, we can simplify the previous expression to
\begin{equation}
    \begin{pmatrix}
    L(z)\\
    R(z)
    \end{pmatrix}
    = e^{\frac{i}{s}\mathcal{Q}_0 (z-\delta_W)}\mathcal{P}\left(e^{\frac{i}{s}\int_{0}^{\delta_W}\mathcal{Q}(z')dz'}\right)
    \begin{pmatrix}
    L(0)\\
    R(0)
    \end{pmatrix},
    \label{Eq:Final_Solution}
\end{equation}
with the constant matrix
\begin{equation}
    \mathcal{Q}_0\equiv
    \begin{pmatrix}
    E & -\mathcal{M}_{0}\\
    \mathcal{M}_{0}^{\dagger} & -E
    \end{pmatrix},
\end{equation}
where $\mathcal{M}_{0} = \mathcal{M}(z>\delta_{W})$ is the mass matrix in the broken phase.

The reflection coefficient from an incident right-handed neutrino, $N_R$, to a left-handed one, $\nu_L$ $(N_L)$, can be simply obtained by imposing a suitable boundary condition. Namely, that at $z=\delta_W$ all states travel towards the inside of the bubble (see Fig.~\ref{Fig:Picture_ref}). 
The matrix $\mathcal{Q}_0$ needs to be diagonalized to find the basis of propagating states inside the bubble so as to set this boundary condition. 
This can be done through the following series of transformations:
\begin{equation}
    \begin{pmatrix}
    \mathcal{U}_L^{\dagger} & 0\\
    0 & \mathcal{V}_R^{\dagger}
    \end{pmatrix}
    \begin{pmatrix}
    E&-\mathcal{M}_{0}\\
    \mathcal{M}_{0}^{\dagger} & -E
    \end{pmatrix}
    \begin{pmatrix}
    \mathcal{U}_L & 0\\
    0 & \mathcal{V}_R
    \end{pmatrix}=
    \begin{pmatrix}
    E & -\mathcal{M}_d\\
    \mathcal{M}_d^T & -E
    \end{pmatrix},
\end{equation}
where $\mathcal{U}_L$ ($\mathcal{V}_R$) is a unitary matrix which diagonalizes $\mathcal{M}_{0}\mathcal{M}_{0}^{\dagger}$ $(\mathcal{M}_{0}^{\dagger}\mathcal{M}_{0})$ such that $\mathcal{M}_d=\begin{pmatrix} 0 & M_d\end{pmatrix}^T$ and $M_d$ is the diagonal $3\times 3$ matrix with the heavy Dirac neutrino mass eigenvalues. Finally, we can do a second transformation $\mathcal{W}$ to rotate $\mathcal{Q}_0$ to its diagonal form
\begin{equation}
    \mathcal{W}^{-1}
    \begin{pmatrix}
    E & 0 & 0 \\
    0 & E & -M_d \\
    0 & M_d & -E
    \end{pmatrix}
    \mathcal{W}
    =
    \begin{pmatrix}
    E & 0 & 0\\
    0 & \sqrt{E^2-M_d^2} & 0\\
    0 & 0 & -\sqrt{E^2-M_d^2}
    \end{pmatrix},
\end{equation}
with
\begin{equation}
    \mathcal{W}\equiv
    \begin{pmatrix}
    1 & 0 & 0\\
    0 & \cosh{\Theta} & \sinh{\Theta}\\
    0 & \sinh{\Theta} & \cosh{\Theta}
    \end{pmatrix},
    \quad
    \tanh{2\Theta}=E^{-1}M_d.
\end{equation}

Performing these rotations, we can now impose the boundary condition at $z=\delta_{W}$ and obtain the reflection coefficient, $\mathcal{R}^u$, as $L(0)=\mathcal{R}^u R(0)$. 
The results for antiparticles ($\overline{\mathcal{R}}^u$) are found by replacing $\mathcal{M}\rightarrow\mathcal{M}^*$ and $\mathcal{U}_L\,(\mathcal{V}_R)\rightarrow \mathcal{U}_L^*\,(\mathcal{V}_R^*)$.
Following a similar procedure and setting the appropriate boundary condition\footnote{Namely, that at $z=0$ there are no states propagating towards the broken phase.}, we can also calculate the transmission coefficient from a state inside the bubble to a left-handed neutrino in the unbroken phase, $\mathcal{T}^b$.
Note that the superscript on the reflection and transmission coefficients denotes the position of the initial state in those processes, either the unbroken phase (``$u$'') or the broken phase (``$b$'').

Now we can calculate the CP asymmetry generated by reflections or transmissions induced in the SM neutrino sector as 
\begin{equation}
    \Delta \mathcal{R}^u(N_{Ri}\rightarrow \nu_{L\alpha})\equiv |\mathcal{R}^u_{\alpha i}|^2-|\overline{\mathcal{R}}^u_{\alpha i}|^2,\quad \Delta \mathcal{T}^b(\mathpzc{N}_i\rightarrow\nu_{L\alpha})\equiv|\mathcal{T}^b_{\alpha i}|^2-|\overline{\mathcal{T}}^b_{\alpha i}|^2,
    \label{Eq:CP_asym}
\end{equation}
where $\mathpzc{N}_i\equiv \begin{pmatrix} \nu_i & N_i\end{pmatrix}^T$ is a propagation eigenstate (either massless, $\nu_i$, or massive, $N_i$) inside the bubble which travels from the broken to the unbroken phase. 
\begin{figure}
    \centering
    \includegraphics[width=0.49\textwidth]{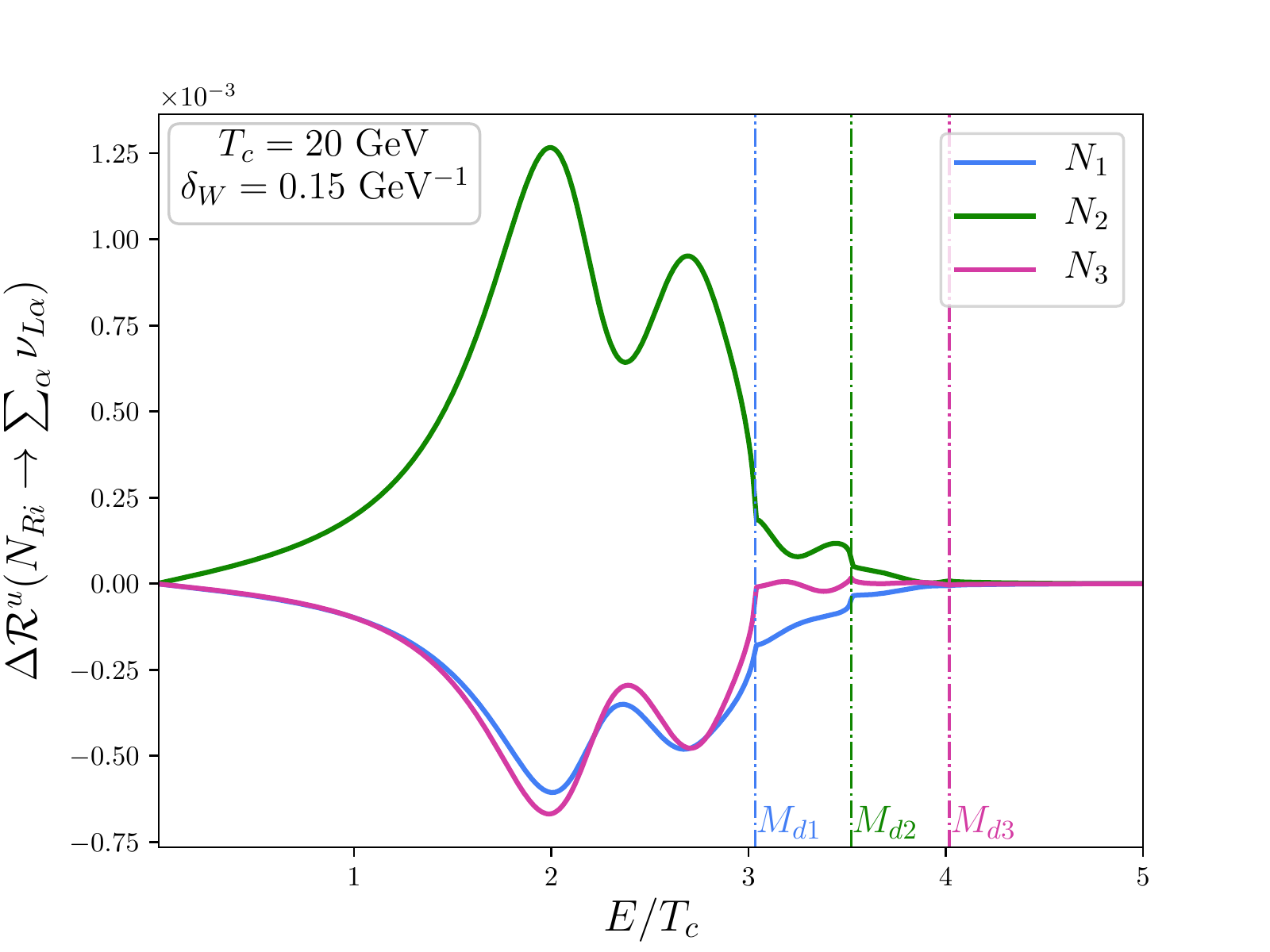}
    \includegraphics[width=0.49\textwidth]{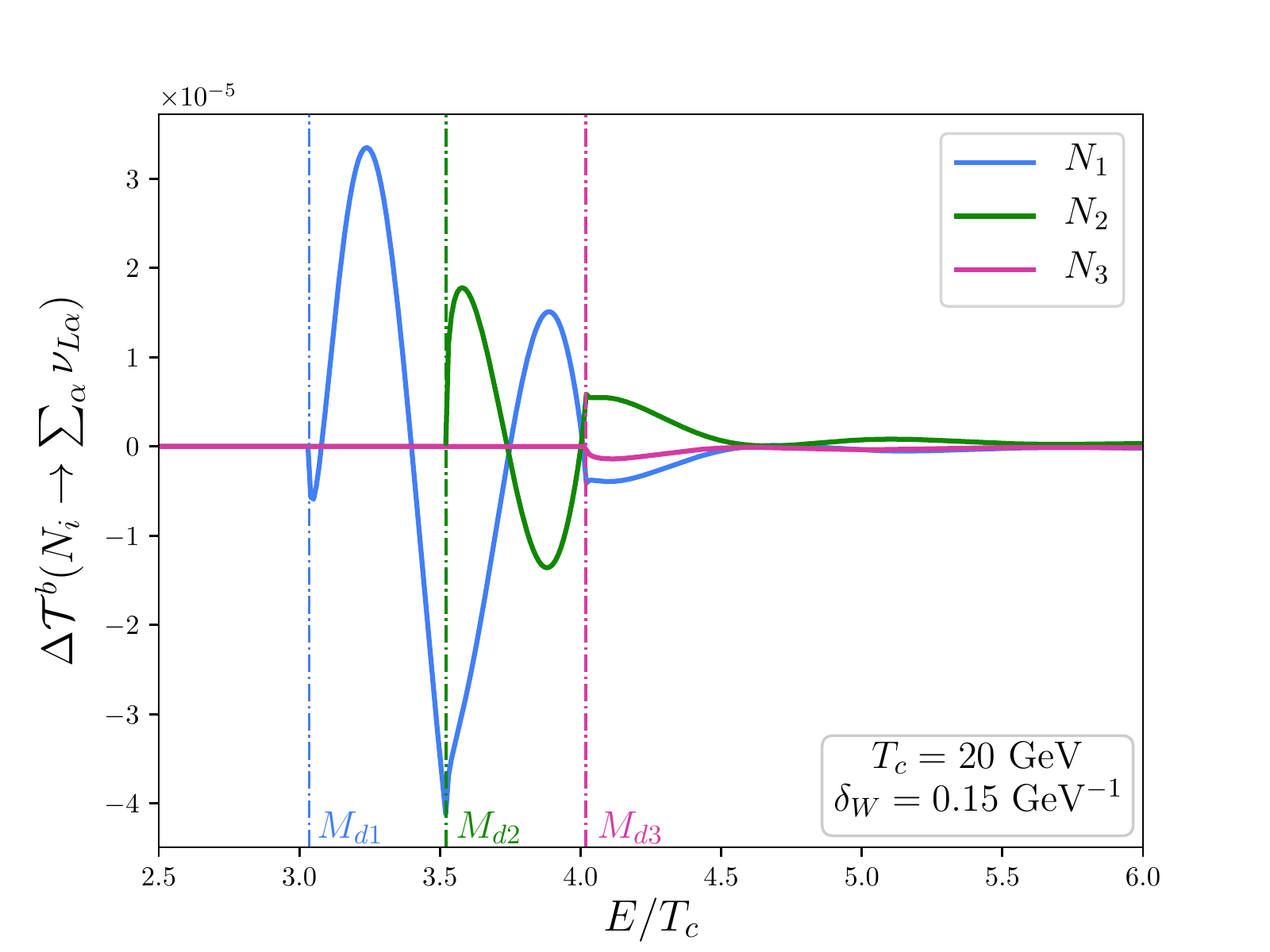}
    \caption{CP asymmetries from the reflection of states from the unbroken phase off the bubble wall (left panel) and transmission of states from the broken to the symmetric phase (right panel). Here we have used the FLOR profile defined in Eq.~(\ref{Eq:Profile_APN}). $M_1$, $M_2$ and $M_3$ are set to 60, 70 and 80~GeV respectively while $m_D$ has been fixed as discussed at the end of Section~\ref{Sec:Lagrangian}. The energy dependence has been normalized to $T_c$, here chosen to be $20$~GeV, but it has no actual impact on the computation of the transmission and reflection coefficients. The heavy-active mixing has been set to $Tr[\theta\theta
   ^\dagger]=0.045$.}
    \label{Fig:CP_asymmetries}
\end{figure}
As an example, the CP asymmetries both for reflection and transmission to $\nu_L$ are presented in Fig.~\ref{Fig:CP_asymmetries} for a benchmark point. Notice that the reflection from $N_R$ to $\nu_L$ is possible for any energy (see left panel), while the transmission from massive states to $\nu_L$ is only possible when the energy is larger than its mass (see right panel). Moreover, as the mass threshold is overcome, the reflection and transmission from massless states are suppressed.

Following Ref.~\cite{Huet:1995sh}, performing an expansion of the $z$-ordered product in Eq.~(\ref{Eq:Final_Solution}), it can be shown that the reflection asymmetry for a particular $\nu_{L\alpha}$ to first non-trivial order goes like
\begin{equation}
    \sum_i\Delta\mathcal{R}^u\left(N_{R i}\rightarrow\nu_{L\alpha}\right)\sim \int_z\sum_{i,j,\beta} f(z) m_{d_\alpha}^2 \text{Im}(V_{Ri \alpha} V_{Ri \beta}^* V_{Rj \alpha}^* V_{Rj \beta}),
    \label{Eq:Approximate_Reflection}
\end{equation}
where $m_{d_\alpha}$ are the eigenvalues of $m_D$ in the broken phase, $f$ is a function of the masses which depends on the position $z$ and the last term is the Jarlskog rephasing invariant defined in Eq.~(\ref{Eq:JarlskogJ}). When summing over all final neutrino states $\nu_{L\alpha}$, the expected Glashow–Iliopoulos– Maiani (GIM) suppression with the differences of the squared masses is found.

Although our main interest is the asymmetry generated in $\nu_L$ as they are charged under $SU(2)_L$ and therefore source sphaleron processes, asymmetries in $N_L$ and $N_R$ are also generated through this mechanism and they may play a relevant role in the BAU generation as we will discuss in the next sections.

\subsection{Vacuum expectation value profiles}\label{Vev_profile}
The dependence of the mass matrix $\mathcal{M}(z)$ on the position $z$ comes solely from the change of the scalar vevs along the bubble wall. It is important to note from Eq.~(\ref{Eq:Final_Solution}) that the Higgs vev, $v_H(z)$, and the one from the singlet scalar, $v_{\phi}(z)$, need to have a different spatial dependence in order for the particle and antiparticle rates to be different. 
Otherwise, one could rotate $\mathcal{Q}(z)$ everywhere to the basis where it is diagonal, finding $\overline{\mathcal{R}}^u=(\mathcal{R}^u)^*$, such that $\Delta \mathcal{R}^u=0$, and the same applies to the transmission coefficients. 
In particular, the authors in Ref.~\cite{Hernandez:1996bu} made the following choice
\begin{equation}
    \frac{v_H(z)}{v_H}=\left\{\begin{array}{llc}
    0, & z\leq0\\
    \frac{z}{\delta_W}, & 0< z\leq \delta_W
    \\1, & z > \delta_W
    \end{array}\right. ,
    \quad
    \quad \frac{v_{\phi}(z)}{v_\phi} =\mathpzc{H}(z),
    \label{Eq:Profile_PN}
\end{equation}
where $\mathpzc{H}(z)$ is the Heaviside step function. We refer to this choice as the Hern\'andez-Rius (HR) profile. In Eq.~(\ref{Eq:Profile_PN}) $v_H\sim246$~GeV and $v_\phi$ are the vevs for the Higgs and singlet scalar in the broken phase, respectively. In the following we will explore $v_{\phi}$ in the range between 2 and 10 TeV. Thus, barring a strong hierarchy among the scalar potential quartic couplings whose study is beyond the scope of this work, we expect the mixing between the higgs and the scalar singlet to be below 10\%, in agreement with present LHC constraints~\cite{Sirunyan:2018owy}. 
It is important to notice that different wall profiles will result in different values of the heavy-active mixing $\theta$ inside the bubble wall and translate to very different sizes for the CP invariant given in Eq.~(\ref{Eq:Jarlskog_Inv}) along the bubble wall. In fact, even though the relative size between $v_{H}(z)$ and $v_{\phi}(z)$ changes within the bubble wall, the HR profile is rather conservative and tends to produce a small CP asymmetry because the mixing $\theta$ in the wall is always smaller than that at the broken phase where strong constrains apply~\cite{Fernandez-Martinez:2016lgt}.

\begin{figure}
    \centering
    \includegraphics[width=0.49\textwidth]{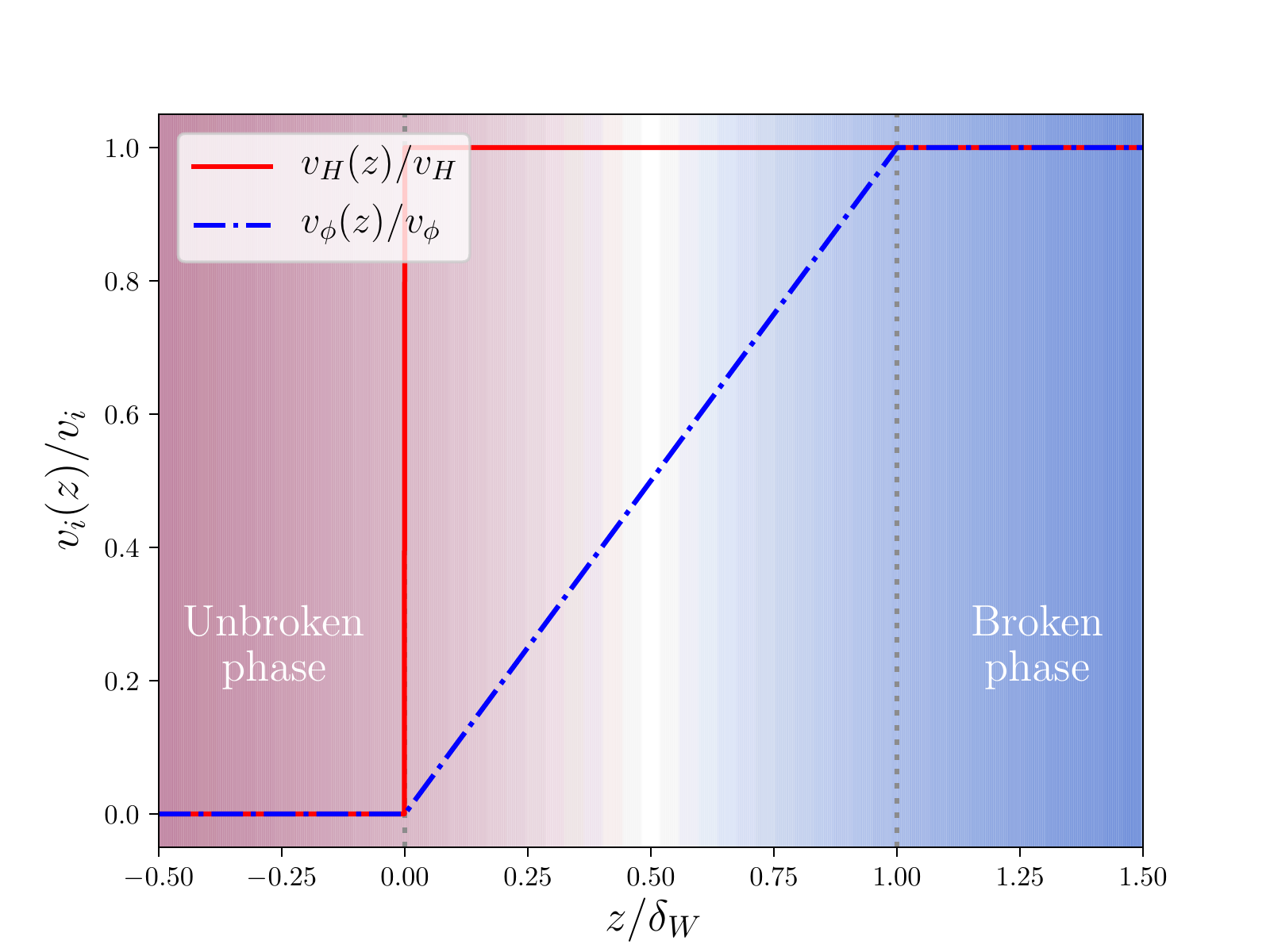}
    \includegraphics[width=0.49\textwidth]{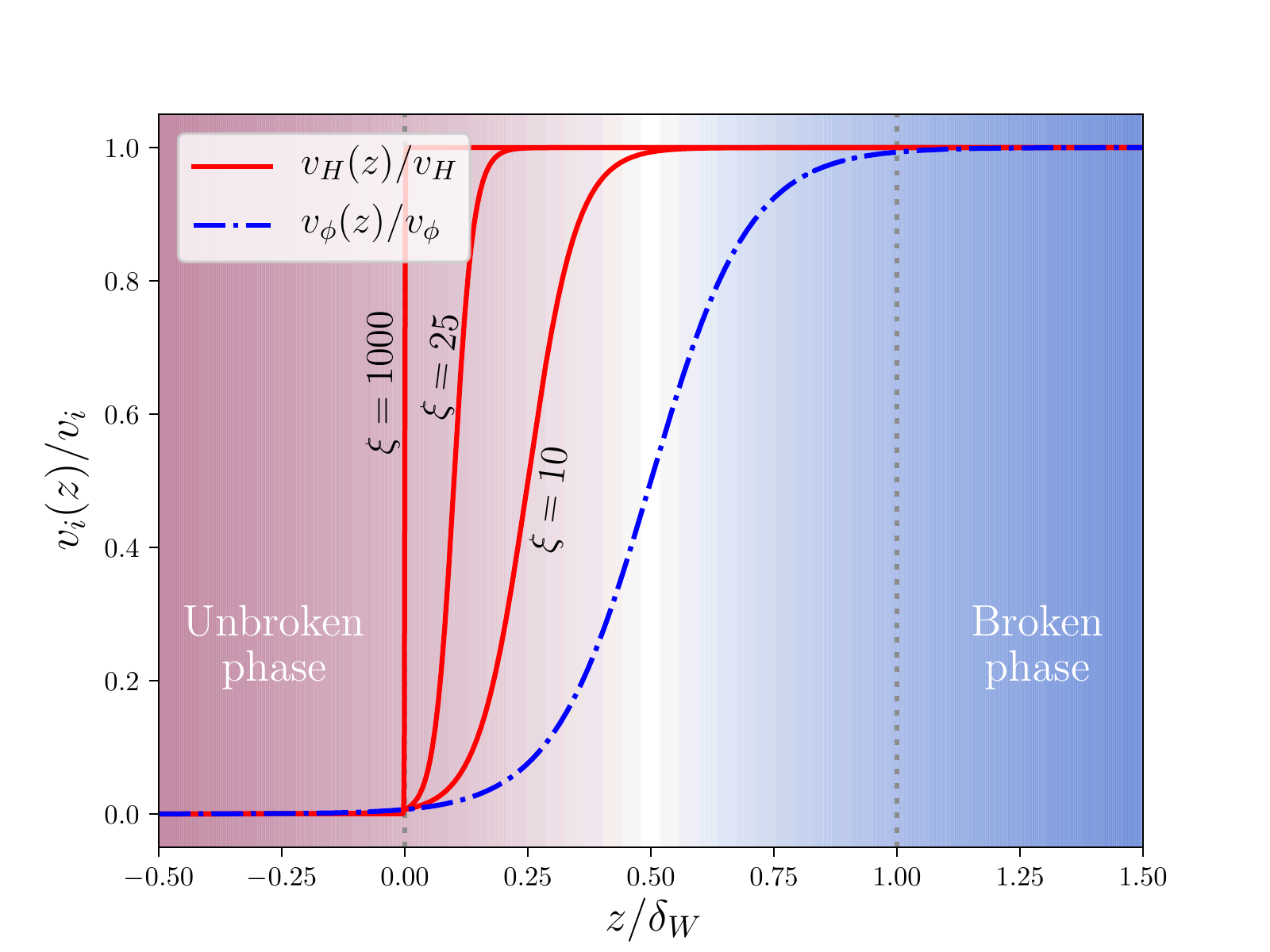}
    \caption{Profile for the vev of the scalars in the bubble wall. The left panel shows the FLOR profile from Eq.~(\ref{Eq:Profile_APN}) while the right panel corresponds to the second set of profiles from Eq.~(\ref{Eq:Profile_Tanh}) with smooth and continuous functions for different values of $\xi$.} 
    \label{Fig:Vev_Profiles}
\end{figure}

We have thus gone beyond Ref.~\cite{Hernandez:1996bu} and studied two particular sets of profiles, which are depicted in Fig.~\ref{Fig:Vev_Profiles}. The first one follows Ref.~\cite{Hernandez:1996bu}, but assigning the profiles to the opposite scalars so as to have larger mixing $\theta$ inside the wall with respect to the broken phase, namely
\begin{equation}
    \frac{v_H(z)}{v_H}=\mathpzc{H}(z),\quad \frac{v_\phi(z)}{v_\phi} =\left\{\begin{array}{llc}
    0, & z\leq0\\
    \frac{z}{\delta_W}, & 0< z\leq \delta_W
    \\1, & z > \delta_W
    \end{array}\right. .
    \label{Eq:Profile_APN}
\end{equation}
We dub this choice as the ``Fern\'andez-L\'opez-Ota-Rosauro'' (FLOR) profile and we will investigate it in detail in the following sections.
The second profile we have studied is a smoothed variant of the FLOR profile, parametrizing the dependence on $z$ with hyperbolic tangents:
\begin{equation}
    \frac{v_H(z)}{v_H}=\frac{1}{2}\left[1+\tanh{\left(\xi \frac{z-\left(5/\xi\right) \delta_W/2}{\delta_W}\right)}\right],\quad \frac{v_{\phi}(z)}{v_{\phi}}=\frac{1}{2}\left[1+\tanh{\left(5\frac{z-\delta_W/2}{\delta_W}\right)}\right].
    \label{Eq:Profile_Tanh}
\end{equation}
We have checked that the particular realization for the profiles from Eq.~(\ref{Eq:Profile_Tanh}), although slightly reducing the final BAU, gives very similar results to the FLOR profile. We will study the dependence of the generated BAU on the $\xi$ parameter, controlling the steepness of the profile for the Higgs vev, at the end of Section~\ref{Sec:Results} and leave an in-depth study of the scalar potential and the vev profiles for future work.

\section{Diffusion equations}\label{Sec:Diff_eq}
The subsequent evolution of the CP asymmetry generated by the interactions with the bubble walls and its eventual conversion into a baryon number asymmetry will be governed by the diffusion equations of the different particle species. 
In principle, all particle species and possible interactions between them should be taken into account, but there are some approximations that can help to simplify the description and make the problem more tractable while providing a good estimation of the baryon asymmetry generated. A fully detailed study of the diffusion equations is beyond the scope of this work, and thus we limit our discussion to two simplified cases which nonetheless contain the relevant physical ingredients, closely following the analysis of Ref.~\cite{Joyce:1994zn}. 

The first case we study contains the most minimal set of diffusion equations,
where we only follow the total baryon and lepton number densities, neglecting all possible wash-out effects and tracking the conversion from lepton to baryon number via the weak sphaleron processes that provide the necessary baryon number violation. We will refer to this as the vanilla scenario, which was studied in Ref.~\cite{Hernandez:1996bu} for the HR profile.

\begin{figure}
    \centering
    \includegraphics[width=0.49\textwidth]{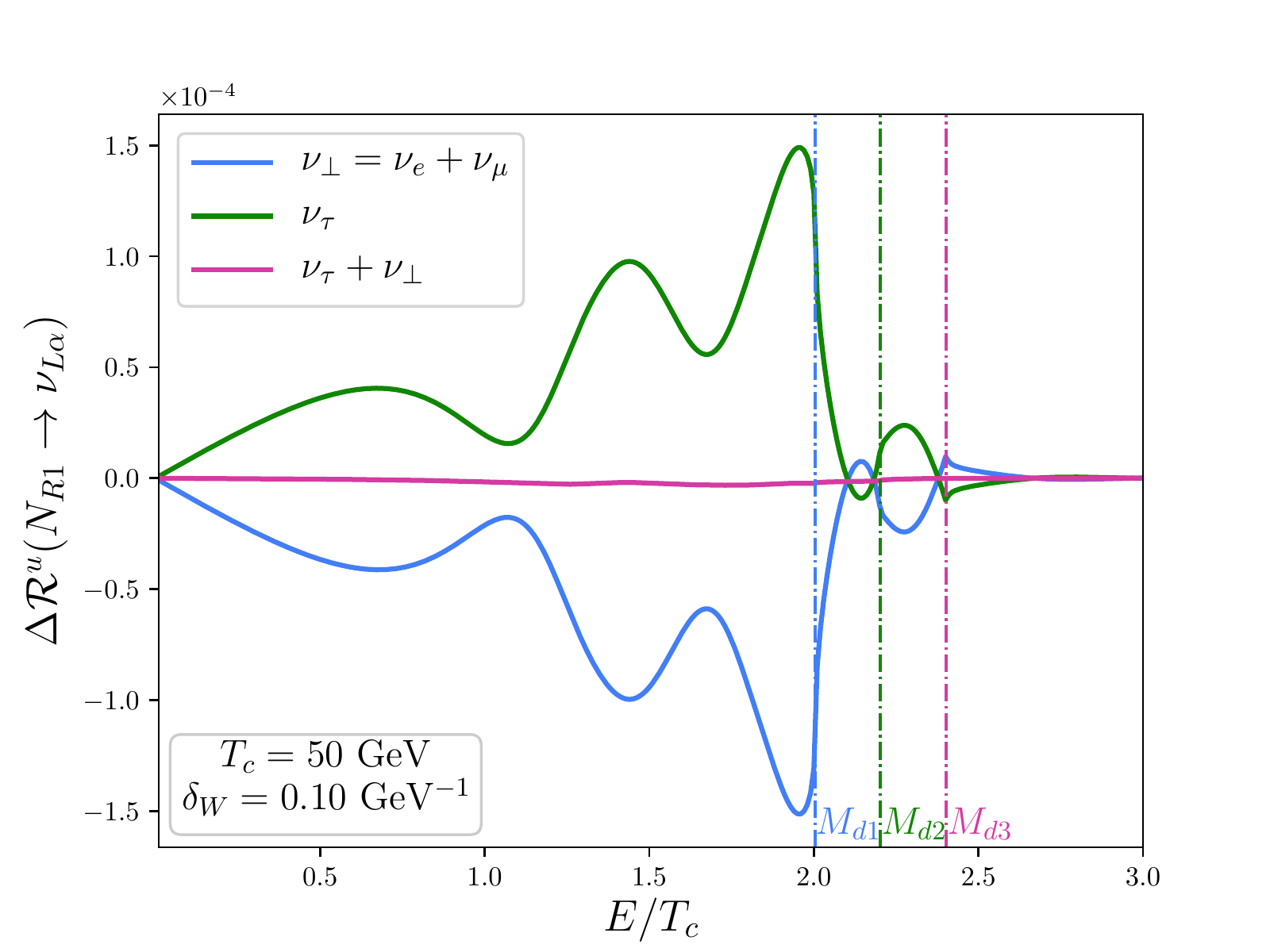}
    \includegraphics[width=0.49\textwidth]{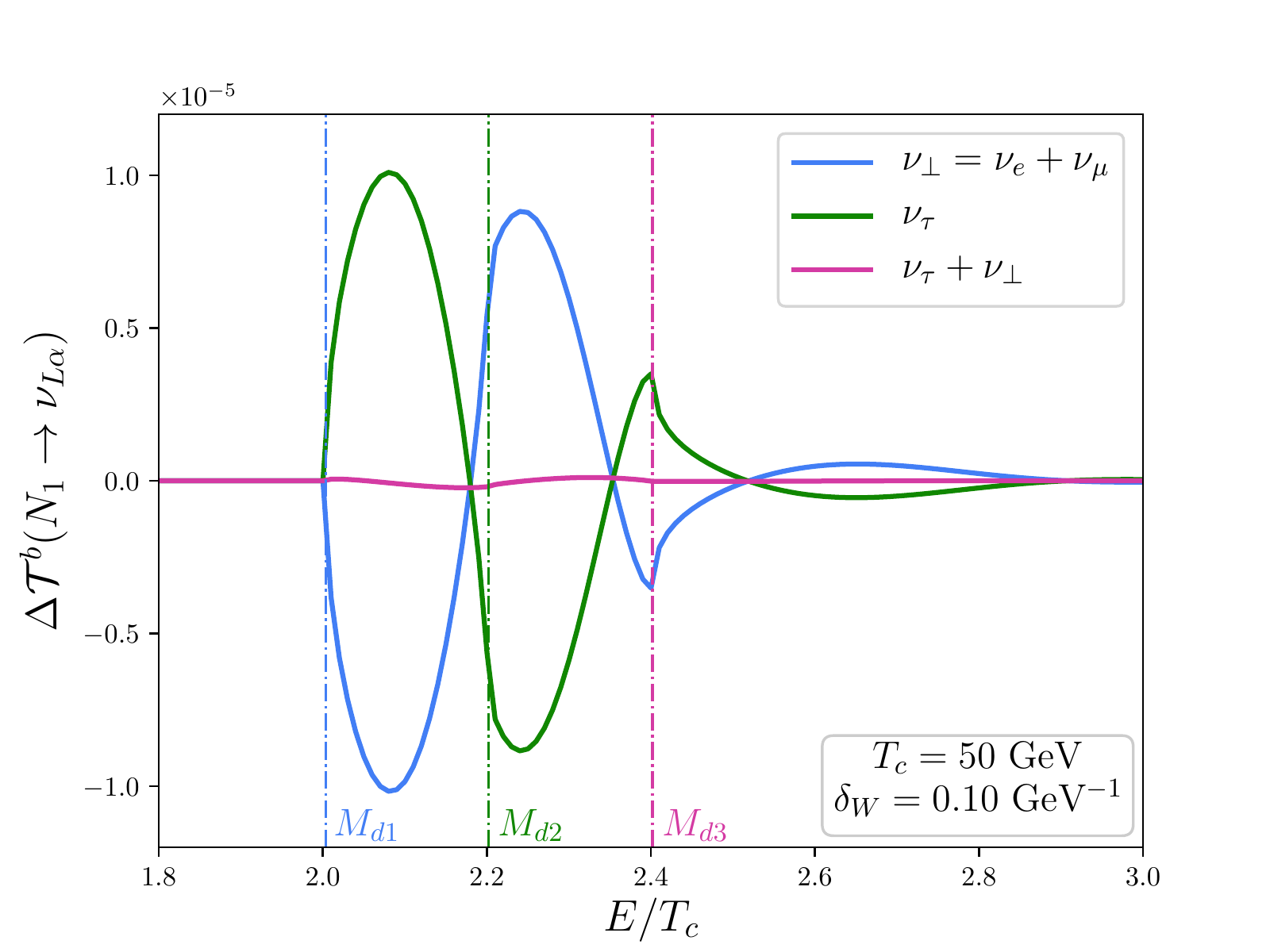}
    \caption{CP asymmetries in the reflection and transmission of $N_{R1}$ (left panel) and $N_1$ (right panel), respectively, into the $\tau$-flavour neutrino or the combination of $e+\mu$ flavours. In magenta we have the sum of the two asymmetries, which tend to cancel each other. Here we have used the FLOR profile defined in Eq.~(\ref{Eq:Profile_APN}). $M_1$, $M_2$ and $M_3$ are set to 100, 110 and 120~GeV respectively while $m_D$ has been fixed as discussed at the end of Section~\ref{Sec:Lagrangian}. The heavy-active mixing has been set to $Tr[\theta\theta^\dagger]=0.007$~\cite{Fernandez-Martinez:2016lgt}. The energy dependence has been normalized to $T_c$, here chosen to be $50$~GeV, but it has no actual impact on the computation of the transmission and reflection coefficients.}
    \label{Fig:CP_asymmetries_flavour}
\end{figure}

In the second case, we introduce the effect of partial wash-out of the asymmetry generated in the different flavours as a further refinement.
In particular, we include the wash-out from the Yukawa interaction between SM and RH neutrinos, which, for some regions of parameter space, may dominate over the sphaleron rate~\cite{Joyce:1994zn}. 
In this case, we will need to follow the asymmetries in the different neutrino species separately, which can prevent the strong cancellation which appears in the total CP asymmetry when summing over all flavour contributions. 
This cancellation among the different flavour contributions originates from the GIM mechanism, as outlined in  Eq.~(\ref{Eq:Approximate_Reflection}). It is depicted both for the reflection and transmission coefficients in an example shown in Fig.~\ref{Fig:CP_asymmetries_flavour}, where we plot separately the CP asymmetry stored in $\nu_\tau$ and $\nu_\perp\equiv\nu_e+\nu_\mu$ as well as the total asymmetry. As can be seen from the figure, the total asymmetry is strongly suppressed as a consequence of the cancellation between the two contributions with the different flavours, which could be prevented through the flavour-dependent wash-out effect. We will refer to this case as the flavoured scenario. 
\subsection{Vanilla scenario}
The minimal set of the diffusion equations we consider is~\cite{Hernandez:1996bu}
\begin{equation}
    \begin{split}
        &D_B \partial_z^2n_B-v_W \partial_zn_B-3\Gamma_S\mathpzc{H}(-z) n_B-\Gamma_S\mathpzc{H}(-z) n_L = 0,\\
        &D_L \partial_z^2n_L-v_W \partial_zn_L-\Gamma_S\mathpzc{H}(-z) n_L-3\Gamma_S\mathpzc{H}(-z) n_B = \xi_L j_{\nu}\partial_z\delta(z),
    \end{split}
    \label{Eq:Diff_Vanilla}
\end{equation}
where we only follow the evolution of total baryon ($n_B$) and lepton number ($n_L$) asymmetries and their conversion through weak sphaleron processes. In Eq.~(\ref{Eq:Diff_Vanilla}), $D_{B(L)}$ is the diffusion constant for baryons (leptons) which we estimate following Ref.~\cite{Joyce:1994zn}, $v_W$ is the wall velocity, $\Gamma_S = 9\kappa \alpha_W^5 T$ is the sphaleron rate with $\kappa\simeq 18$~\cite{DOnofrio:2014rug} and $\alpha_W$ the weak coupling constant and we have neglected the bubble width. The CP current generated through reflections and transmissions of neutrinos, $j_{\nu}$, can be computed from the coefficients derived in the previous sections convoluted with the corresponding distribution functions for each species:

\begin{equation}
    \begin{split}
    j_{\nu}= \frac{1}{\gamma}\sum_{i,\alpha}\int \frac{d^3p}{(2\pi)^3}\bigg\{&\Delta \mathcal{T}^b(\mathpzc{N}_i\rightarrow \nu_{L\alpha})\frac{|p^b_{zi}|}{E^b_i}f^b_i(p^b_{i})+
    \Delta\mathcal{R}^u(N_{Ri}\rightarrow\nu_{L\alpha})\frac{|p^u_{zi}|}{E^u_i}f^u_i(p^u_{i})\bigg\},
    \end{split}
    \label{Eq:j_nu_Vanilla}
\end{equation}
where $p^b_{zi} \in (-\infty, 0]$ and $p^u_{zi} \in [0, \infty)$ are the momentum perpendicular to the bubble wall for transmissions and reflections, respectively. 
The gamma factor $\gamma\equiv 1/\sqrt{1-v_{W}^{2}}$ comes from boosting to the wall rest frame where $\Delta \mathcal{T}^b$ and $\Delta\mathcal{R}^u$ are computed. 
The energy of the particle $i$ in the broken phase is defined as $E_i^b \equiv \sqrt{p_T^2+(p^b_{zi})^2+m_i^2}$ with $p_T$ the transverse momentum and $m_i$ the physical mass of the particle, while the energy in the unbroken phase is given by $E_{i}^{u} = \sqrt{p_{T}^{2} + (p^{u}_{z i})^{2}}$, since all the particles are massless.
The distribution function $f^{b (u)}_{i}$ for an initial state with index $i$ in the broken (unbroken) phase is the Fermi-Dirac distribution boosted to the wall rest frame:
\begin{equation}
    f^b_i(p_i^b)=\frac{1}{1+e^{\frac{\gamma}{T}(E_i^b-v_W p_{zi}^b)}},\quad f^u_i(p_i^u)=\frac{1}{1+e^{\frac{\gamma}{T}(E_i^u-v_W p_{zi}^u)}}.
\end{equation}

Finally, $\xi_L$ parametrizes the persistence length of the current in the vicinity of the wall and is estimated in Ref.~\cite{Joyce:1994zn} as $\xi_L\sim 6 D_L v_i$, where $v_i$ is the mean velocity of the reflected and transmitted particles. In the following, we will conservatively estimate the generated BAU assuming $\xi_L\sim D_L$, although our survey over the points of interest in the parameter space shows that $v_i\sim 0.6-0.8$ and we do not obtain any values below $0.4$. 

\begin{figure}
    \centering
    \includegraphics[width=0.65\textwidth]{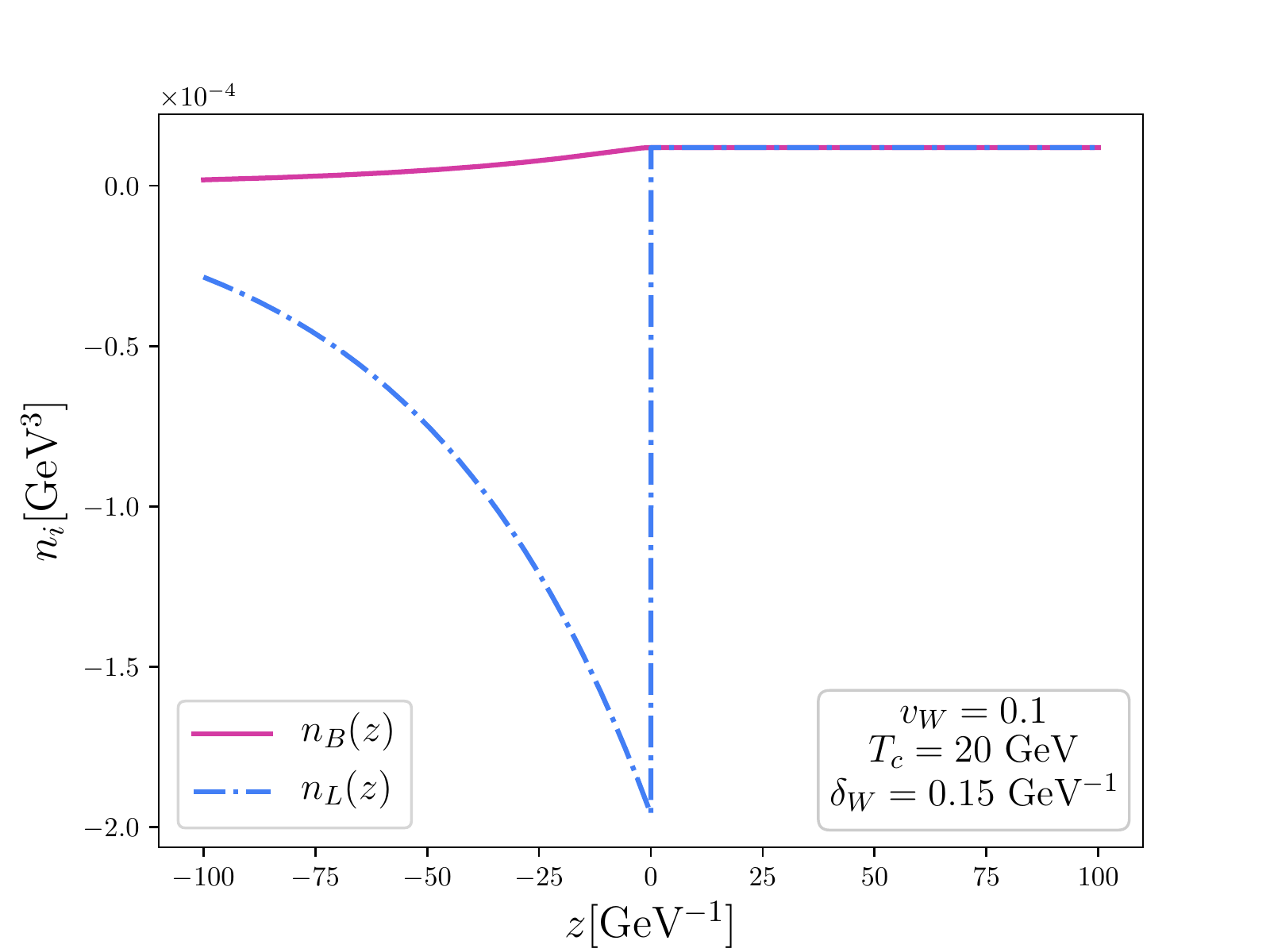}
    \caption{Evolution of lepton and baryon number asymmetries as they transition from the symmetric phase ($z\rightarrow -\infty$) to the broken phase ($z\rightarrow \infty$). The masses of the heavy neutrinos in this case are $M_1= 60$~GeV, $M_2= 70$~GeV and $M_3= 80$~GeV while $m_D$ has been fixed as discussed at the end of Section~\ref{Sec:Lagrangian}. The heavy-active mixing has been set to $Tr[\theta\theta^\dagger]=0.045$. The final BAU found for this particular point is $Y_B\sim Y_B^{obs}/3$.}
    \label{Fig:Number_density_Vanilla}
\end{figure}

In order to solve Eq.~(\ref{Eq:Diff_Vanilla}), a set of boundary conditions needs to be imposed. In particular, the asymmetry in the number densities should vanish far from the wall when $z\rightarrow-\infty$ and become constant as $z\rightarrow \infty$ inside the broken phase.
Additionally, by integrating once and twice the diffusion equations given in Eq.~(\ref{Eq:Diff_Vanilla}), we find the following continuity equations along the bubble wall:
\begin{equation}
    D_i \partial_z n_i -v_W n_i\big|_-^+ = 0,\; \text{for}\;i=B,L,
    \label{Eq:Cont_1}
\end{equation}
and
\begin{equation}
    D_B n_B\big|_-^+=0,\quad D_L n_L\big|_-^+=\xi_L j_{\nu},
    \label{Eq:Cont_2}
\end{equation}
respectively. This means that the lepton number density presents a discontinuity between $z=0^-$ and $z=0^+$ due to the injected CP asymmetry in the SM neutrinos. 
The solutions of the diffusion equations Eq.~\eqref{Eq:Diff_Vanilla} can be expressed as
\begin{equation}
    \begin{split}
        &n_B=\left\{\begin{array}{lcc}
        B_1e^{k_1 z}+B_2e^{k_2z}, & z<0\\
        B, & z>0
        \end{array}\right.,\quad
        n_L=\left\{\begin{array}{lcc}
        L_1e^{k_1 z}+L_2e^{k_2z}, & z<0\\
        L, & z>0
        \end{array}\right.,
    \end{split}
\end{equation}
where $k_i>0$ are the solutions to the following cubic equation:
\begin{equation}
    D_B D_L k^3-v_W(D_B+D_L)k^2+\left[v_w^2-\Gamma_S(3D_L+D_B)\right]k+4v_W\Gamma_S=0.
    \label{Eq:Cubic_Valilla}
\end{equation}
The constants $B_{1,2}$, $L_{1,2}$, $B$, and $L$ are determined using Eq.~(\ref{Eq:Diff_Vanilla}), (\ref{Eq:Cont_1}), and (\ref{Eq:Cont_2}). $B$ corresponds to the baryon number asymmetry in the broken phase, which we find to be
\begin{equation}
    B=\frac{\Gamma_S v_W \xi_L j_{\nu}}{D_L^2 k_1 k_2(D_B k_1+D_B k_2-v_W)},
    \label{Eq:B_Vanilla}
\end{equation}
where we observe, as expected, that the baryon number is proportional to the injected lepton asymmetry and that the proportionality constant depends on the sphaleron rate, the expansion velocity of the bubble wall, and the diffusion of particles in the symmetric phase.
The final asymmetry, $Y_{B}$, will be given by
\begin{equation}
    Y_B = \frac{B}{s(T_c)},
\end{equation}
where $s(T_c)$ is the entropy density at the temperature $T_c$.
A solution to the diffusion equations can be found in Fig.~\ref{Fig:Number_density_Vanilla} for a benchmark parameter point. 
In Fig.~\ref{Fig:Number_density_Vanilla} we can see how a baryon asymmetry $n_B$ is slowly generated approaching the bubble wall ($z\rightarrow0$) and is then frozen out at a given value inside the bubble ($z\rightarrow\infty$) where sphalerons are no longer effective.

\subsection{Flavoured scenario}\label{SubSec:Flavour}
A potentially relevant effect not considered in Eq.~(\ref{Eq:Diff_Vanilla}) are the (flavour-dependent) wash-out processes. The most important contribution from the SM charged leptons would be that of the tau Yukawa coupling to the Higgs boson. Nonetheless, the rate of this interaction is still smaller than the sphaleron rate~\cite{Joyce:1994zn}:
\begin{equation}
    \frac{\Gamma_\tau}{T}\sim 0.28\alpha_W Y_\tau^2\ll \frac{\Gamma_S}{T}=9\kappa \alpha_W^5,
\end{equation}
where $Y_\tau\sim 0.01$ is the SM tau Yukawa coupling. Thus, we will neglect these contributions. 

However, the neutrino Yukawa couplings may also mediate wash-out processes and are naturally sizable in the low-scale Seesaw scenarios assumed in this work. In particular, the SM neutrino flavour eigenstates may annihilate with the massless $N_{Ri}$ in the symmetric phase to third generation quarks via an s-channel Higgs exchange, washing out the CP asymmetry stored in $\nu_{L\alpha}$. 
The rate for these interactions is given by~\cite{Joyce:1994zn}
\begin{equation}
    \frac{\Gamma_{N_{Ri}\nu_{L\alpha}}}{T}\sim\frac{1}{128\pi}\left(Y_t^2+Y_b^2\right)|(Y_{\nu})_{\alpha i}|^2\sim 0.0024|\theta_{\alpha i}|^2\frac{2M_{i}^2}{v_H^2},
    \label{Eq:Neutrino_Yukawa_rate}
\end{equation}
where $Y_t(Y_b)$ is the top (bottom) Yukawa coupling, and we have replaced the neutrino Yukawa coupling with the heavy-active neutrino mixing $\theta_{\alpha i}$ defined in Eq.~\eqref{Eq:Mixing} and the Dirac mass $M_{i}$ of a heavy neutrino $N_{R i}$.
From Eq.~(\ref{Eq:Neutrino_Yukawa_rate}) and given the present bounds on the mixing~\cite{Fernandez-Martinez:2016lgt} at $2\sigma$, it is possible to have $\Gamma_{N_{Ri}\nu_{L\alpha}}>\Gamma_S$ for heavy Dirac neutrinos with masses $M_{i}\gtrsim 200$~GeV. 
Therefore, the possible wash-out between $\nu_{L\alpha} \leftrightarrow N_{Ri}$ can, in principle, play an important role in the parameter regions with large $M_{N}$.
Furthermore, a sizable CP invariant from Eq.~(\ref{Eq:Jarlskog_Inv}) requires some hierarchy in the mixing $\theta$, making it thus necessary to consider the different flavours with different wash-out rates. 
Taking the wash-out effect due to $Y_{\nu}$ into account,
the extended set of flavour-dependent diffusion equations we consider is the following
\begin{equation}
    \begin{split}
    &D_B\partial_z^2n_B-v_W\partial_zn_B-3\Gamma_S\mathpzc{H}(-z)n_B-\Gamma_S\mathpzc{H}(-z)\left(n_{\nu_e}+n_{\nu_\tau}\right)=0,\\
    &D_L\partial_z^2n_{\nu_e}-v_W\partial_zn_{\nu_e}-3\Gamma_S\mathpzc{H}(-z)n_B-\Gamma_S\mathpzc{H}(-z)\left(n_{\nu_e}+n_{\nu_\tau}\right)\\
    &\quad -\Gamma_{N_1\nu_e}\left(\frac{1}{2}n_{\nu_e}-n_{N_1}\right)-\Gamma_{N_2\nu_e}\left(\frac{1}{2}n_{\nu_e}-n_{N_2}\right)=\xi_L j_{\nu_e}\partial_z\delta(z),\\
    &D_L\partial_z^2n_{\nu_\tau}-v_w\partial_zn_{\nu_\tau}-3\Gamma_S\mathpzc{H}(-z)n_B-\Gamma_S\mathpzc{H}(-z)\left(n_{\nu_e}+n_{\nu_\tau}\right)\\
    &\quad -\Gamma_{N_1\nu_\tau}\left(\frac{1}{2}n_{\nu_\tau}-n_{N_1}\right)-\Gamma_{N_2\nu_\tau}\left(\frac{1}{2}n_{\nu_\tau}-n_{N_2}\right)=\xi_L j_{\nu_\tau}\partial_z\delta(z),\\
    &D_{R_1}\partial_z^2n_{N_1}-v_W\partial_z n_{N_1}+\Gamma_{N_1\nu_e}\left(\frac{1}{2}n_{\nu_e}-n_{N_1}\right)+\Gamma_{N_1\nu_\tau}\left(\frac{1}{2}n_{\nu_\tau}-n_{N_1}\right)=\xi_{R_1}j_{N_1}\partial_z\delta(z) ,\\
    &D_{R_2}\partial_z^2n_{N_2}-v_W\partial_z n_{N_2}+\Gamma_{N_2\nu_e}\left(\frac{1}{2}n_{\nu_e}-n_{N_2}\right)+\Gamma_{N_2\nu_\tau}\left(\frac{1}{2}n_{\nu_\tau}-n_{N_2}\right)=\xi_{R_2}j_{N_2}\partial_z \delta(z),
    \end{split}
    \label{Eq:Full_Diffusion_Flavour}
\end{equation}
where the current $j_{\nu_{\alpha}}$ of the CP asymmetry in $\nu_{L\alpha}$ is defined as Eq.~\eqref{Eq:j_nu_Vanilla} but without taking the sum over the flavour index $\alpha$.

There are also source terms for the $N_R$, $j_{N i}$, arising from reflections and transmissions, that can be computed similarly to the ones for $\nu_{L}$ and may also become relevant since they are linked to the active neutrino CP asymmetry through the potentially sizable Yukawa interactions of Eq.~(\ref{Eq:Neutrino_Yukawa_rate}). 
We estimate the diffusion constants for the RH neutrinos as 
\begin{equation}
    D_{R_i}^{-1}\sim max\left\{Y_{\nu}^4,Y_N^4\right\}(4\pi)^{-2}T
\end{equation}
following Ref.~\cite{Long:2017rdo}. 
Therefore, if the Yukawa coupling between the heavy neutrinos and the singlet scalar $Y_{N}$ dominates the diffusion constants\footnote{This happens in the parameter space of interest as long as $Y_N\gtrsim0.04$ for the lightest heavy neutrino $N_{R1}$.}, $D_{R_{2(3)}}\sim D_{R_1} M_{N1}^4/M_{N2(3)}^4$. 
Thus, the smallest diffusion constant is the one of the heaviest neutrino $N_{R3}$, while the $N_{R3}$ Yukawa couplings to the SM flavour eigenstates are equal in size to those of $N_{R2}$ (see discussion after Eq.~(\ref{Eq:JarlskogJ})). 
Therefore, the impact of the evolution of $N_{R3}$ in the flavour eigenstates will be smaller than that of $N_{R2}$ and is expected, in any case,
to be between the results of the following two simplified cases.
In the first case we simply neglect its influence altogether, while in the second scenario we overestimate it by assuming that the diffusion constant for $N_{R3}$ is the same as for $N_{R2}$. 
Both limiting cases are conveniently described by the reduced set of Eqs.~(\ref{Eq:Full_Diffusion_Flavour}). In particular, when $N_{R3} $ is assumed to have the same diffusion coefficient as $N_{R2} $ it is only necessary to replace $j_{N_2} \to j_{N_2}+j_{N_3}$ in the source term and $\left( 1/2 n_{\nu_\alpha}-n_{N_2} \right) \to \left(n_{\nu_\alpha}-n_{N_2}\right)$ in the wash-out terms. 
\begin{figure}
    \centering
    \includegraphics[width=0.65\textwidth]{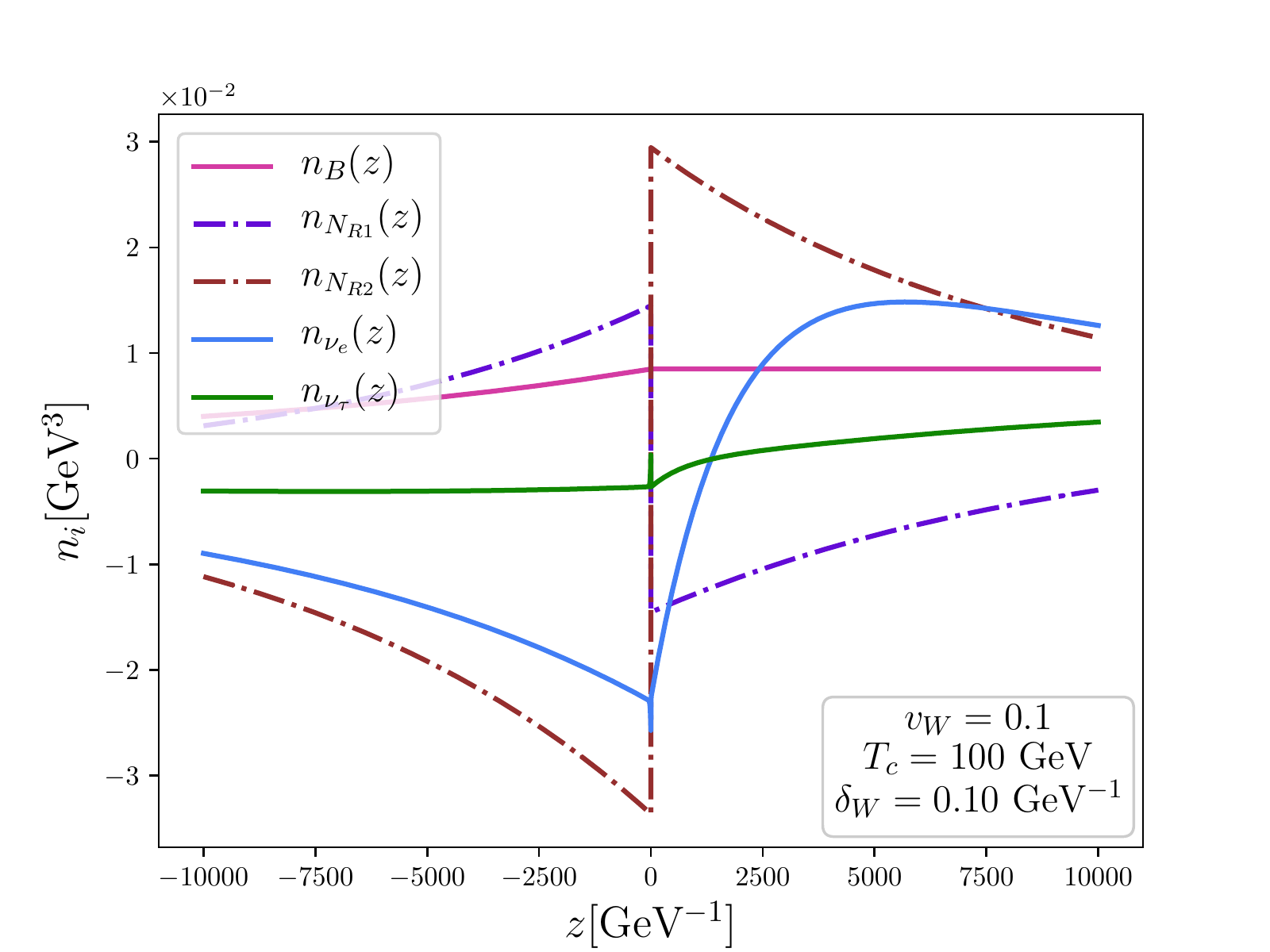}
    \caption{Evolution of the number densities for baryons and $\nu_e$ and $\nu_\tau$ from the symmetric phase ($z\rightarrow -\infty$) to the broken phase ($z\rightarrow \infty$) in the flavoured scenario. The masses of the heavy neutrinos in this particular case are $M_1= 80$~GeV, $M_2= 90$~GeV and $M_3= 160$~GeV while $m_D$ has been fixed as discussed at the end of Section~\ref{Sec:Lagrangian}. The heavy-active mixing has been set to $Tr[\theta\theta^\dagger]=0.007$~\cite{Fernandez-Martinez:2016lgt}. The final BAU found for this particular point is $Y_B\sim 2\times Y_B^{obs}$.}
    \label{Fig:Number_density_Flavoured}
\end{figure}
 An additional equation following the $\nu_\mu$ density has not been considered since, as outlined in the discussion after Eq.~(\ref{Eq:JarlskogJ}), we chose $m_d=diag(m_{d_\tau}/\sqrt{2}, 0, m_{d_\tau})$, and hence no asymmetry is generated in $\nu_\mu$. 
 Notice that the neutrino flavours in Eqs.~(\ref{Eq:Full_Diffusion_Flavour}) are only labels and do not necessarily correspond to the actual electron, $\mu$ or $\tau$ flavours as discussed at the end of Section~\ref{Sec:Lagrangian}. 

In Fig.~\ref{Fig:Number_density_Flavoured} we show the solution of the diffusion equations for the different particle densities for a benchmark parameter point. 
As expected, even though the injected asymmetries for the different neutrino flavours tend to cancel due to the GIM mechanism, the different wash-out rates from interactions with the RH singlet neutrinos partially prevent the cancellation and thus a larger asymmetry than in the vanilla case may be generated.

\section{Results}\label{Sec:Results}
In this section, we parametrize the mass matrix so as to maximize the invariant from Eq.~(\ref{Eq:Jarlskog_Inv2}) as discussed at the end of Section~\ref{Sec:Lagrangian}, leaving only one free parameter, $m_{d_\tau}$, which can be constrained through the bounds on heavy neutrino mixing~\cite{Fernandez-Martinez:2016lgt}. 
\begin{figure}
    \centering
    \includegraphics[width=0.49\textwidth]{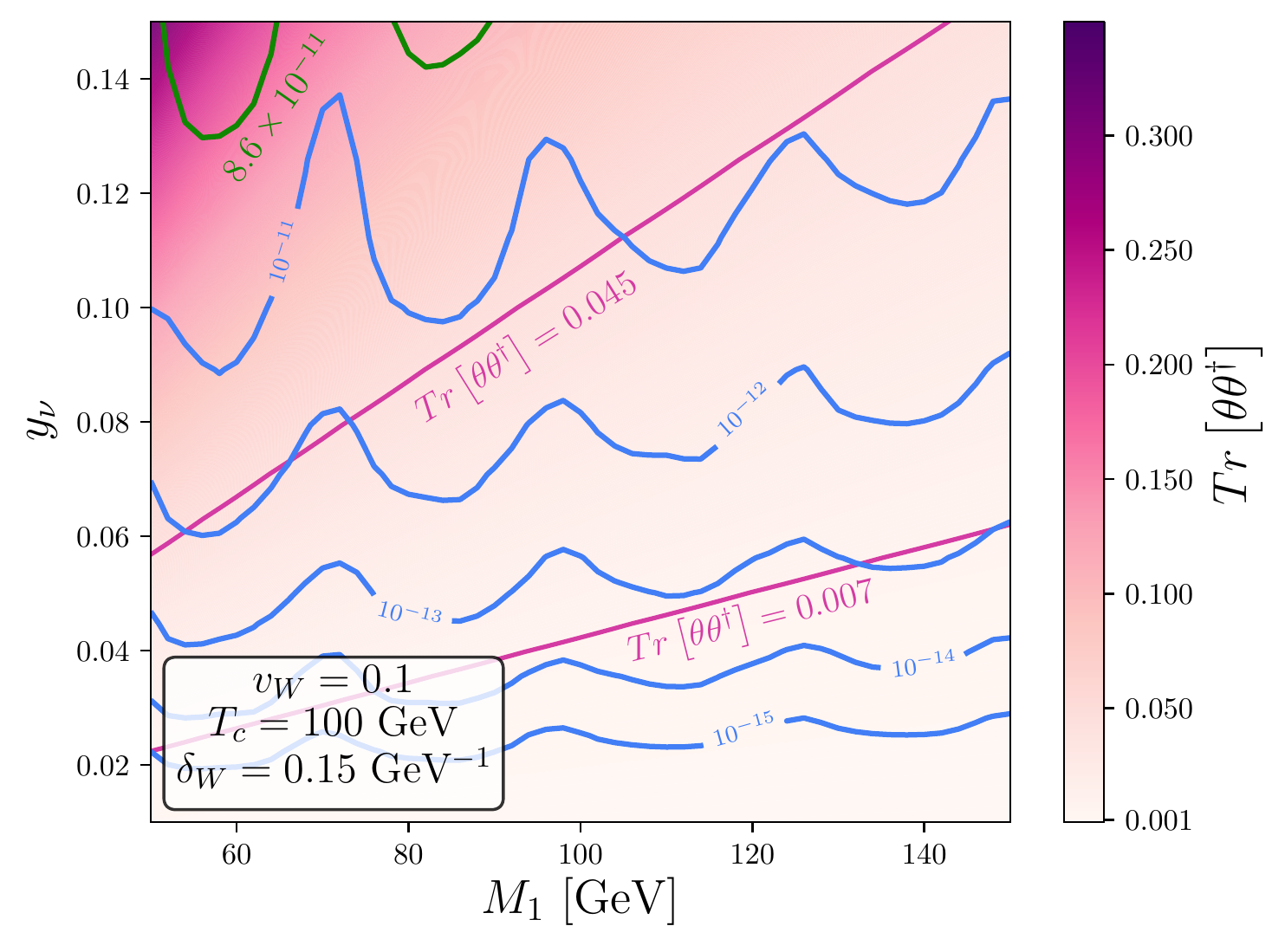}
    \includegraphics[width=0.49\textwidth]{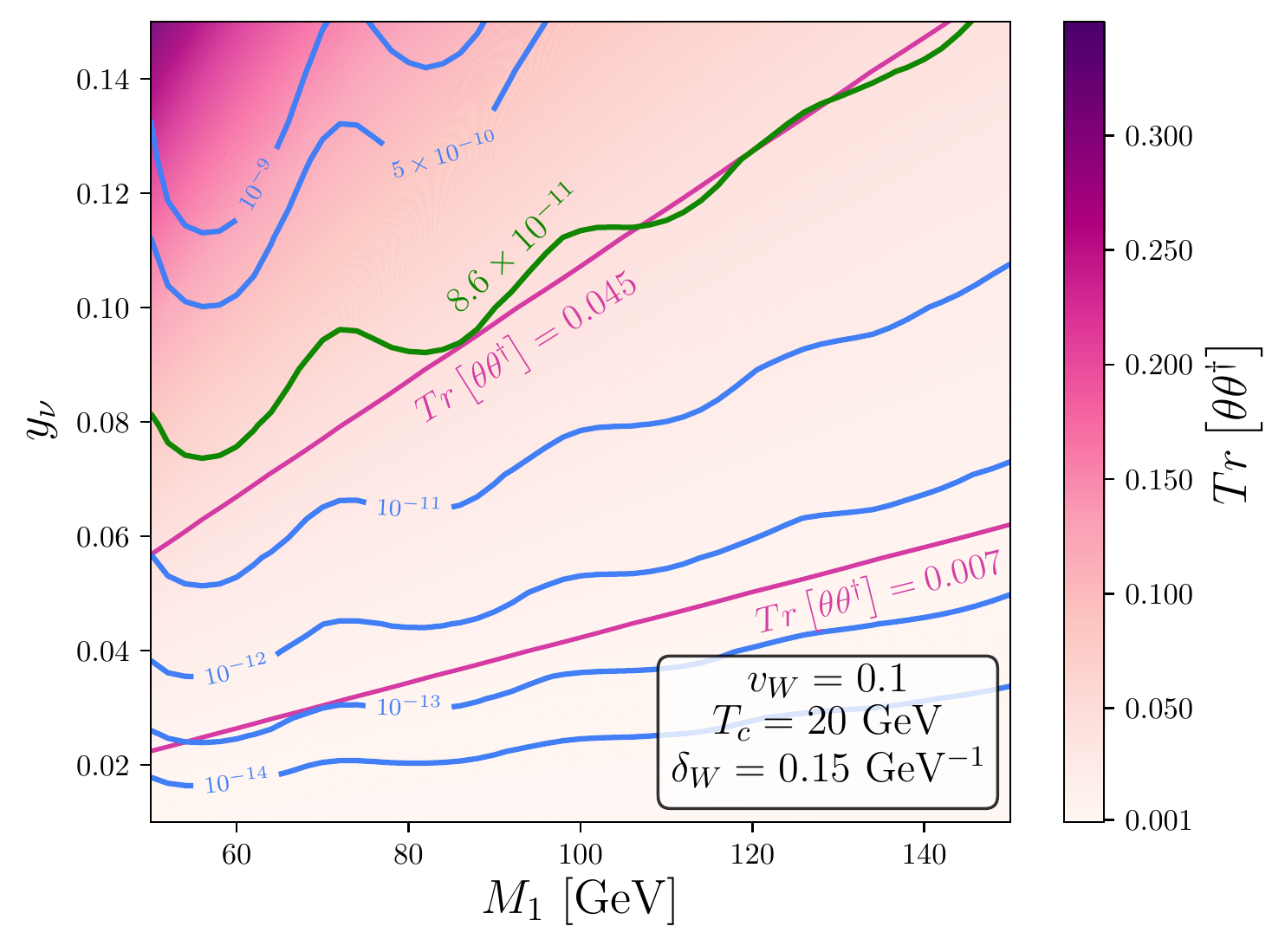}
    \caption{Resulting baryon asymmetry as a function of the Yukawa coupling $y_{\nu}$ and the smallest heavy neutrino mass $M_1$ in the vanilla case for two different temperatures of the phase transition, $T_c=100$~GeV (left panel) and $T_c=20$~GeV  (right panel). The masses of the other two heavy neutrinos are $M_2=M_1+10$~GeV and $M_3=M_2+10$~GeV while $m_D$ has been fixed as discussed at the end of Section~\ref{Sec:Lagrangian}. Along the green line the observed BAU is reproduced. The value of $Tr\left[\theta\theta^{\dagger}\right]$ is represented by the color bar legend, while the current bound for this quantity is represented with the two magenta lines for $Tr\left[\theta\theta^{\dagger}\right]=0.007$~\cite{Fernandez-Martinez:2016lgt} and $Tr\left[\theta\theta^{\dagger}\right]=0.045$ taking into account or not the invisible width of the $Z$, respectively.}
    \label{Fig:Y_B_Mass_vs_Yukawa}
\end{figure}
In Fig.~\ref{Fig:Y_B_Mass_vs_Yukawa} we present contours of constant baryon yield, $Y_B$, generated in the vanilla scenario as a function of the mass of the lightest singlet neutrino $M_1$ and the Yukawa coupling $y_\nu\equiv \sqrt{2}m_{d_\tau}/v_H$, using the FLOR profile. The other two heavy neutrino masses have been fixed to $M_2=M_1+10$~GeV and $M_3=M_2+10$~GeV. We show our results for two cases with different temperatures $T_c$. 
The colour shading indicates the value of the neutrino mixing $Tr[\theta\theta^\dagger]$, while the magenta lines correspond to the $2\sigma$ bounds from electroweak precision and flavour observables including (not including) the invisible decay of the $Z$ boson~\cite{Fernandez-Martinez:2016lgt}: $Tr[\theta\theta^\dagger]=0.007$ ($Tr[\theta\theta^\dagger]=0.045$).

As expected, larger $y_\nu$ and lighter $M_N$ translate into larger heavy-active mixing, enhancing the final CP asymmetry and consequently the final $Y_B$. 
Given the strong constraints from precision electroweak and flavour observables imposing $Tr[\theta\theta^\dagger]\leq0.007$, in both cases the final BAU falls short by two or three orders of magnitude.
Thus, even though the FLOR profile maximizes neutrino mixing along the wall width, the bounds on this mixing today are too stringent to generate the observed baryon asymmetry.
Therefore, we conclude that it is not possible to explain the matter-antimatter asymmetry within the vanilla scenario unless the constraints on the heavy-active neutrino mixing in the broken phase can somehow be evaded.

A possibility in this direction would be that the singlet heavy neutrinos couple to some other dark species, making their decays invisible.
Moreover, for heavy neutrino masses below the mass of the $Z$ boson, $M_Z$, the bounds on the mixing would be relaxed to $Tr[\theta\theta^\dagger]\sim 0.045$ since one of the most stringent constraints, the one stemming from the invisible width of the $Z$, would also be avoided. 
In this case, for low temperatures of the phase transition such as $T_c=20$~GeV, it is indeed possible to generate the observed asymmetry, as can be seen in the right panel of Fig.~\ref{Fig:Y_B_Mass_vs_Yukawa}.
\begin{figure}
    \centering
    \includegraphics[width=0.7\textwidth]{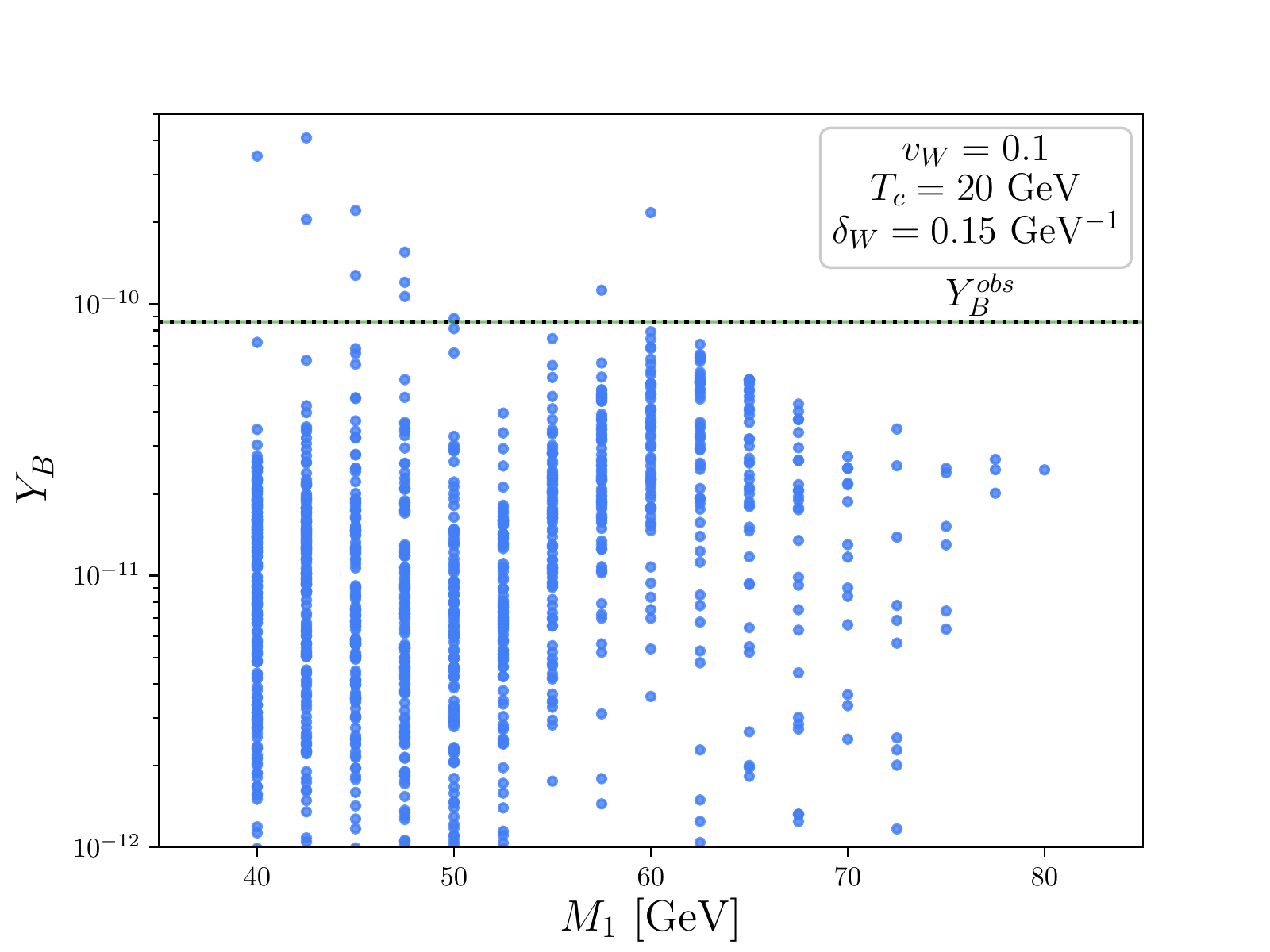}
    \caption{Generated baryon asymmetry if the invisible decay of the $Z$ boson does not apply to the bounds on neutrino mixing. In this case the singlet neutrinos need to be lighter than the $Z$, favoring lower temperatures for the phase transition. The different mass ranges are $M_1\in[40$~GeV$,M_Z]$, $M_2\in[M_1+2.5$~GeV$,M_Z]$ and $M_3\in[M_2+2.5$~GeV$,M_Z]$, scanned in steps of $2.5$~GeV, while $m_D$ has been fixed as discussed at the end of Section~\ref{Sec:Lagrangian}. The heavy-active mixing has been set to $Tr[\theta\theta^\dagger]=0.045$.}
    \label{Fig:BAU_NoZ}
\end{figure}
This is further confirmed in Fig.~\ref{Fig:BAU_NoZ}, where we scan over the three heavy neutrino masses assuming $M_i<M_Z$. As can be observed in the figure, the BAU can be explained in some small regions of parameter space for $T_c=20$~GeV. Finally, to highlight the effect of the vev profile assumed, we scanned over the same parameter space for the HR profile of Ref.~\cite{Hernandez:1996bu}. We find that the BAU generated is typically suppressed by about $2-3$ orders of magnitude with respect to the one obtained with the FLOR profile.

\begin{figure}
    \centering
    \includegraphics[width=0.49\textwidth]{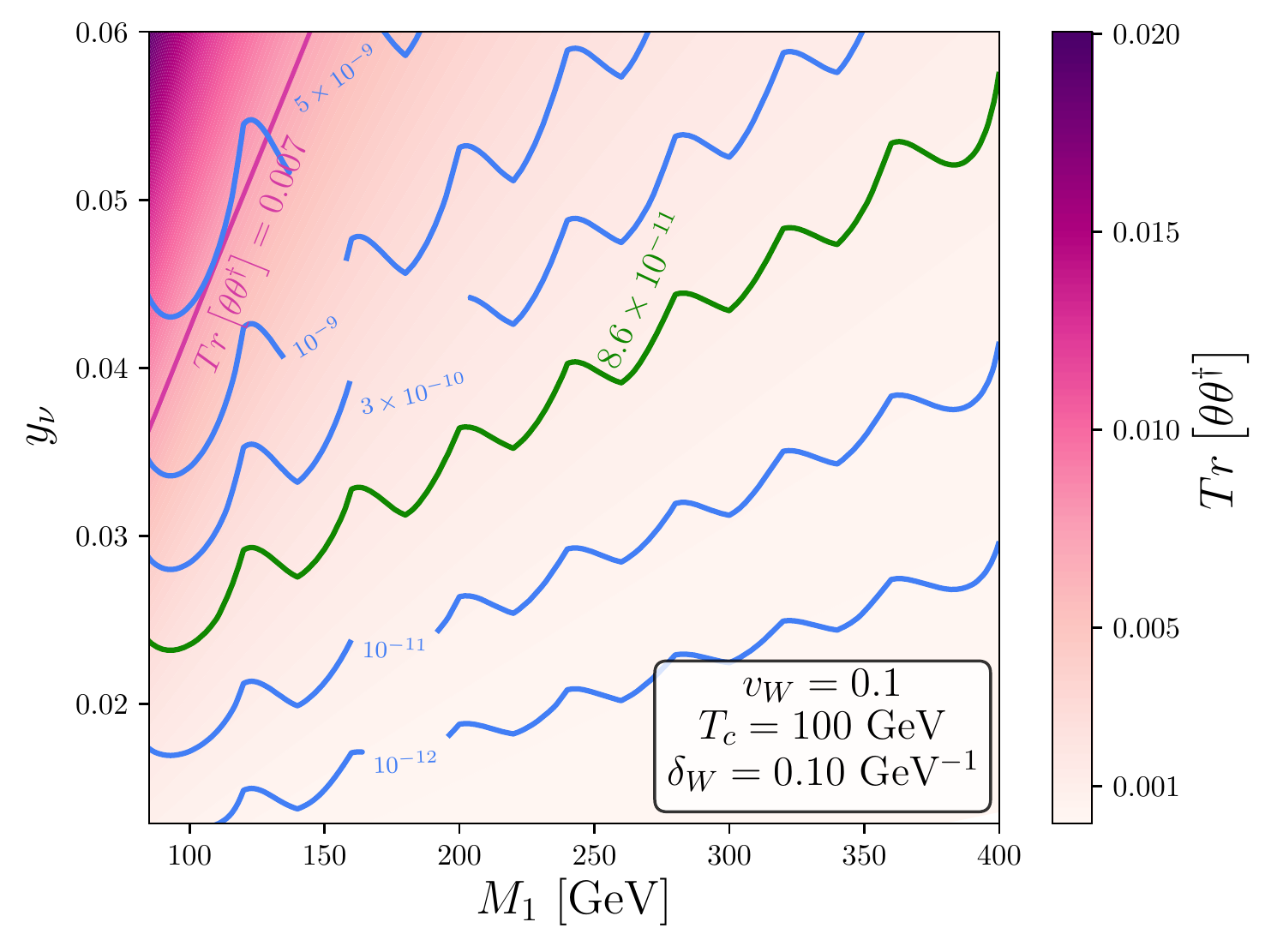}
    \includegraphics[width=0.49\textwidth]{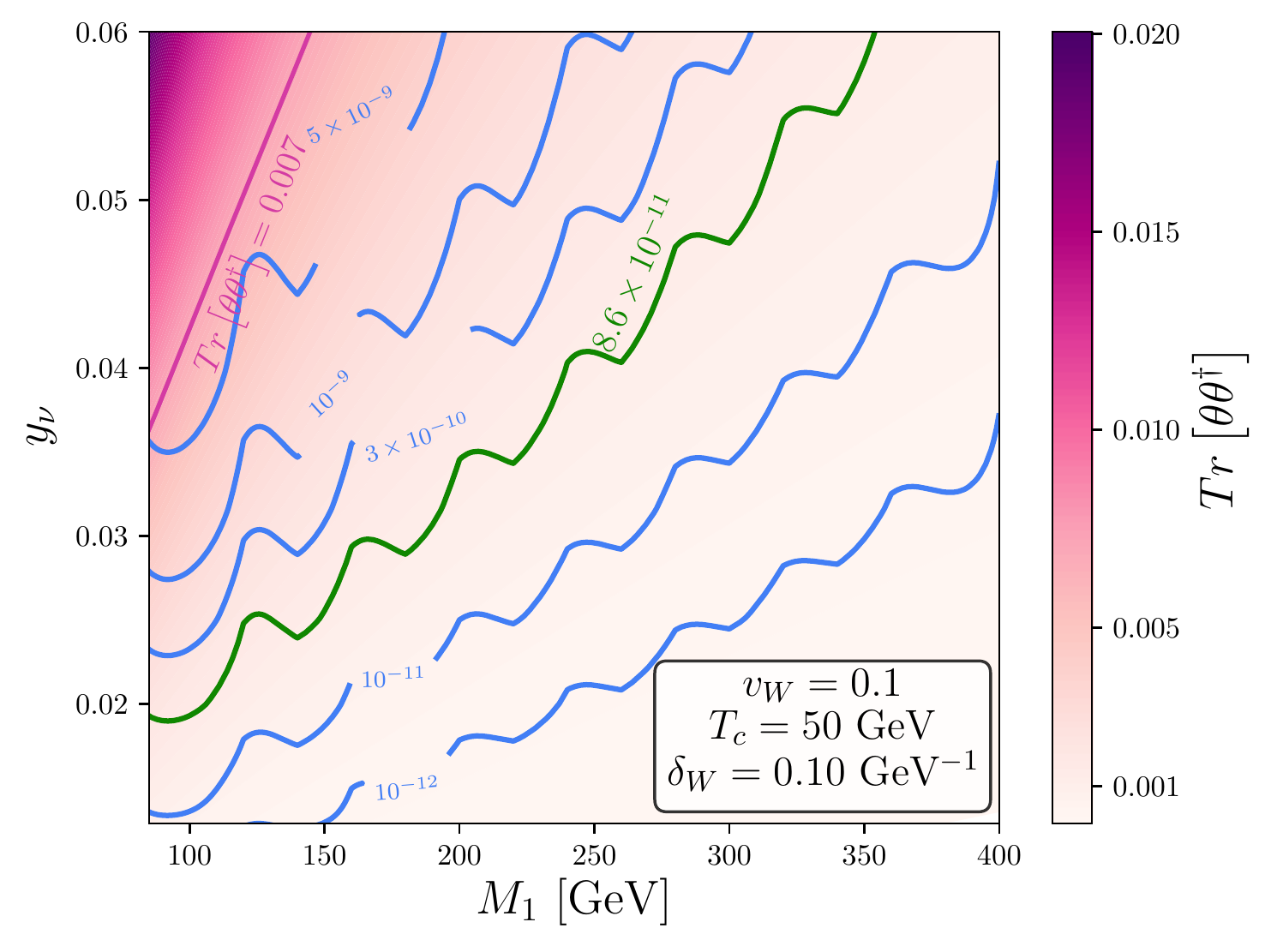}
    \caption{Resulting baryon asymmetry as a function of the Yukawa coupling $y_{\nu}$ and the smallest heavy neutrino mass $M_1$ in the flavoured scenario for $T_c=100$~GeV (left panel) and $T_c=50$~GeV (right panel). The masses of the other two heavy neutrinos are $M_2=M_1+10$~GeV and $M_3=M_2+10$~GeV. Along the green line the observed BAU is reproduced. The value of $Tr\left[\theta\theta^{\dagger}\right]$ is represented by the color bar legend, while the bound for this quantity is represented with the magenta line for $Tr\left[\theta\theta^{\dagger}\right]=0.007$~\cite{Fernandez-Martinez:2016lgt}.}
    \label{Fig:Y_B_Flavour_Mass_Yukawa}
\end{figure}

In Fig.~\ref{Fig:Y_B_Flavour_Mass_Yukawa} we present our results on the final BAU generated in the flavoured scenario where we include the wash-out effect due to the interaction between $\nu_L$ and $N_R$. 
We show contours of $Y_B$ as a function of $M_1$ and $y_{\nu}$ while the colour gradation indicates the value of $Tr[\theta\theta^\dagger]$. The bound on the mixing at $2\sigma$ from electroweak precision and flavour observables is shown as a magenta contour~\cite{Fernandez-Martinez:2016lgt}.
As expected, in contrast to the results for the vanilla scenario shown in Fig.~\ref{Fig:Y_B_Mass_vs_Yukawa}, introducing the flavour effects prevents the GIM cancellation found when summing over different species, and thus the baryon asymmetry can potentially be explained within present bounds. 
Moreover, for the regions of parameter space with some hierarchy in the RH neutrino spectrum and hence in their diffusion constants, the corresponding GIM cancellation in the RH sector asymmetry is also prevented. 
This asymmetry can also be converted into a SM neutrino asymmetry and then to a baryon one through the Yukawa and sphaleron processes, respectively. 
Thus, flavour effects enhance the final baryon asymmetry in a two-fold way, and a baryon asymmetry significantly larger than that in the vanilla scenario is obtained, as shown in Fig.~\ref{Fig:Y_B_Flavour_Mass_Yukawa} (to be compared with Fig.~\ref{Fig:Y_B_Mass_vs_Yukawa}).

\begin{figure}
    \centering
    \includegraphics[width=0.49\textwidth]{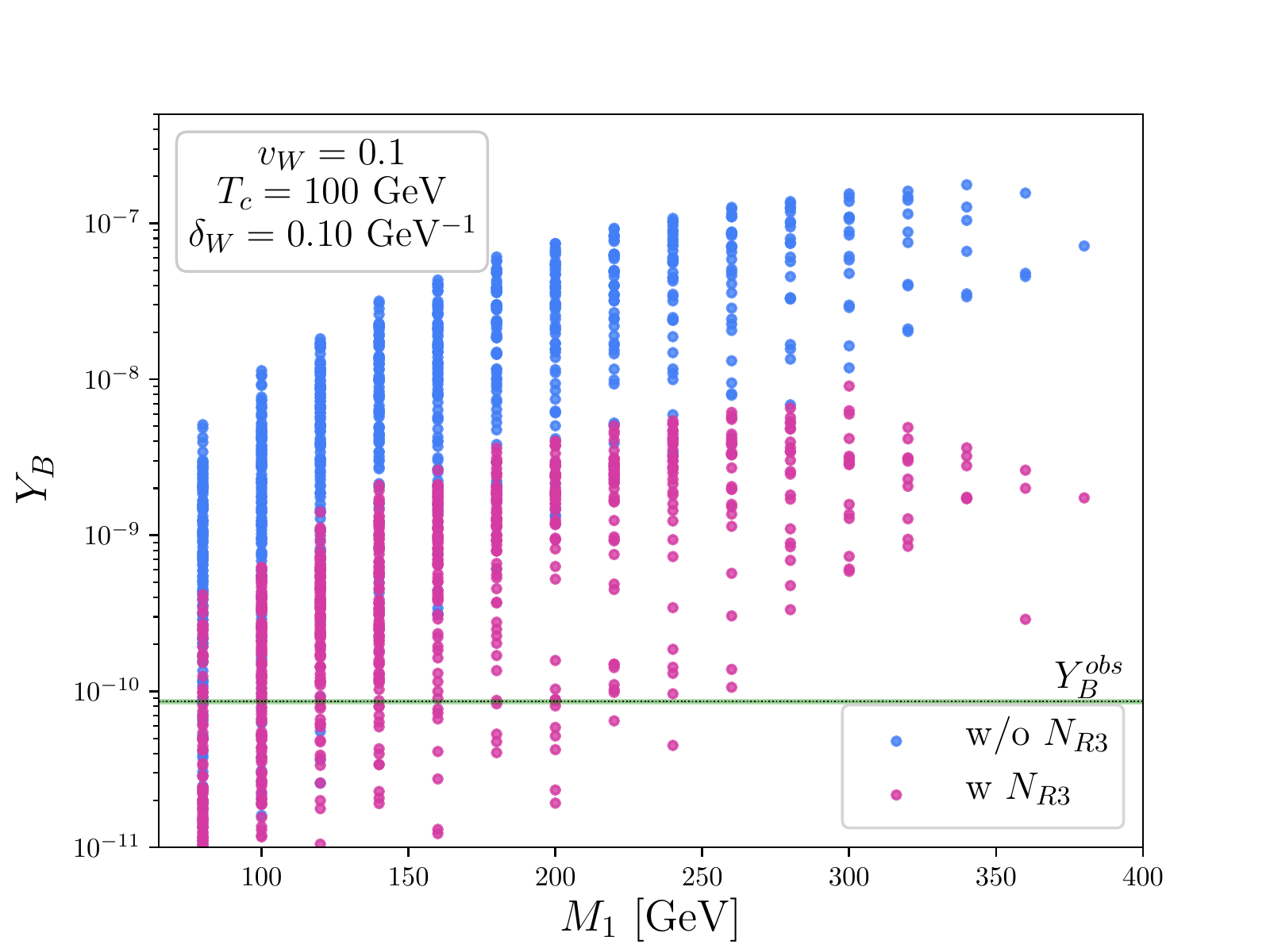}
    \includegraphics[width=0.49\textwidth]{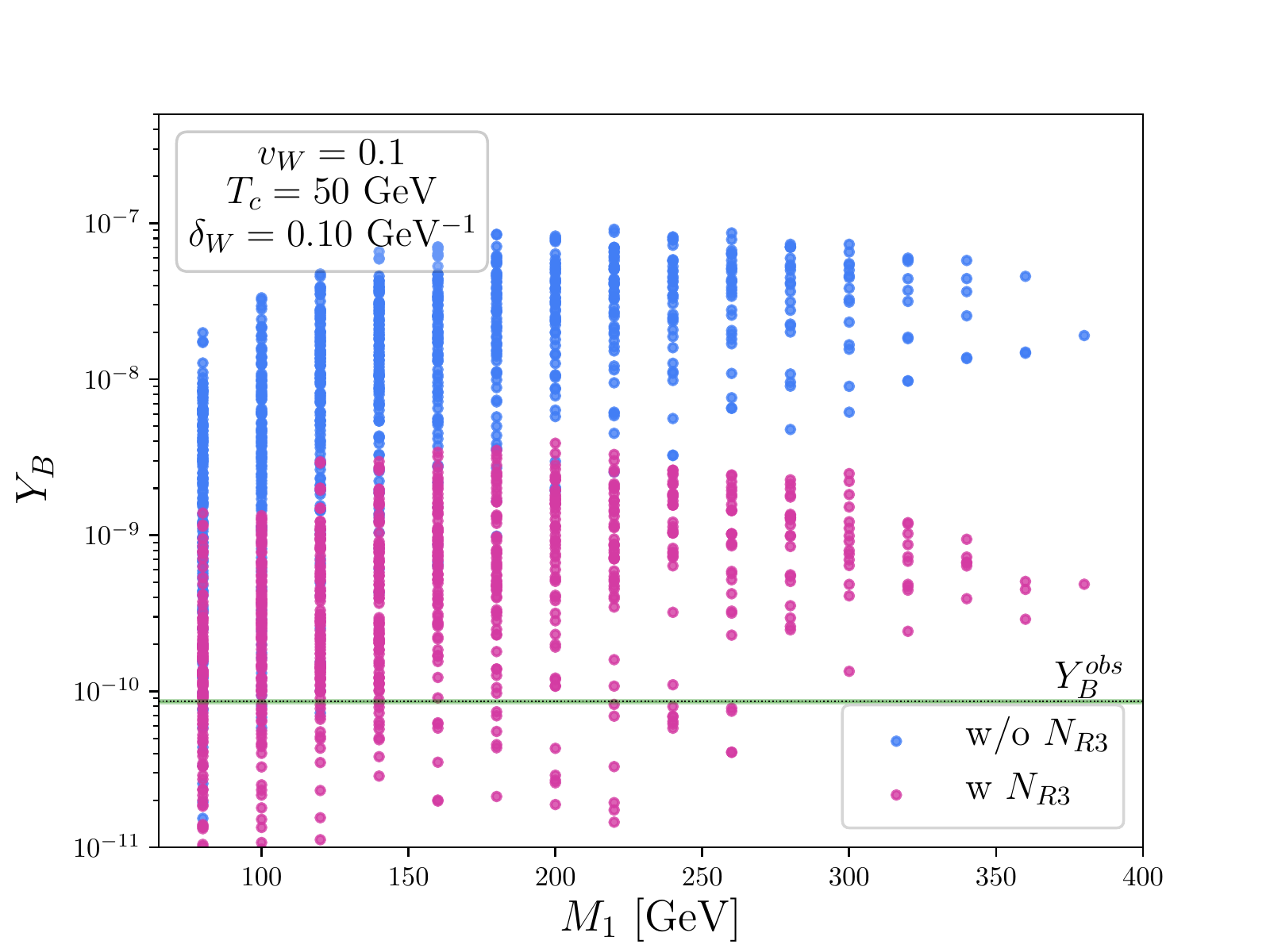}
    \caption{Final BAU generated in the flavoured scenario for different masses of the singlet neutrinos and two temperatures for the phase transition, $T_c=100$~GeV (left panel) and $T_c=50$~GeV (right panel). The mass ranges are $M_1\in[80,400]$~GeV, $M_2\in[M1+10,400]$~GeV and $M_3\in[M2+10,400]$~GeV, scanned in steps of $20$~GeV, while $m_D$ has been fixed as discussed at the end of Section~\ref{Sec:Lagrangian}. For the blue dots the contribution of $N_{R3}$ has been neglected as in Eq.~(\ref{Eq:Full_Diffusion_Flavour}), while the magenta points overestimate its importance, as described at the end of Section~\ref{SubSec:Flavour}. The heavy-active mixing has been set to $Tr[\theta\theta^\dagger]=0.007$~\cite{Fernandez-Martinez:2016lgt}.}
    \label{Fig:BAU_Flavour-T100}
\end{figure}

Indeed, in Fig.~\ref{Fig:BAU_Flavour-T100} in which the three heavy neutrino masses are scanned over a large range of values, we find that most sample points can reproduce or exceed the observed BAU. 
The main contribution to the BAU actually stems from the injection of the particle asymmetry in the $N_R$ sector. Since its diffusion coefficients are much larger because of its weaker interactions, they may more efficiently induce asymmetries in the other species. 
In general, the asymmetry becomes larger for larger $T_c$ because particles in the broken phase suffer from Boltzmann suppression, thus explaining why for $T_c=50$~GeV the asymmetry does not increase for larger $M_1$ while it does for $T_c=100$~GeV. 
As mentioned at the end of Section~\ref{Sec:Diff_eq}, we have also analyzed the case in which $N_{R3}$ is taken into account with its diffusion constant taken to be equal to that of $N_{R2}$, which is an overestimation of its importance.
This is depicted by the magenta dots in Fig.~\ref{Fig:BAU_Flavour-T100}, for which the generated BAU is reduced with respect to the blue dots in which the role of $N_{R3}$ was neglected. 
Indeed, the asymmetries generated in the $N_R$, analogously to the ones generated for $\nu_L$, tend to cancel each other through the GIM mechanism when a sum over all possible states is performed. 
The actual contribution of $N_{R3}$ with its corresponding diffusion constant would yield a result lying between the two limits corresponding to the magenta and blue points. 
Finally, we have estimated the impact of some effects we did not incorporate in our analysis, such as the inclusion of possible decoherence within the bubble wall or of thermal masses, and conclude that, for the parameters studied here, they can induce $\mathcal{O}(1)$ corrections that would not modify our conclusions.

It is interesting to point out that, even though Fig.~\ref{Fig:BAU_Flavour-T100} shows that, within the approximations performed, the present constraints allow for a generation of a BAU up to 2 orders of magnitude larger than the observed one, we choose the neutrino Yukawa couplings so as to maximize the relevant CP invariant throughout this study. 
Moreover, the final BAU scales with three powers of $\theta\theta^\dagger$, as can be seen in Eq.~(\ref{Eq:Jarlskog_Inv}). 
Thus, improving our present constraints on $\theta\theta^\dagger$ by about a factor 5 could potentially allow to probe the whole parameter space for the setup and make it testable at the LHC and future collider experiments~\cite{delAguila:2008cj,Atre:2009rg,Antusch:2015mia,Deppisch:2015qwa,Antusch:2016ejd,Cai:2017mow,Dev:2018kpa,Pascoli:2018heg}. 

\begin{figure}
    \centering
    \includegraphics[width=0.65\textwidth]{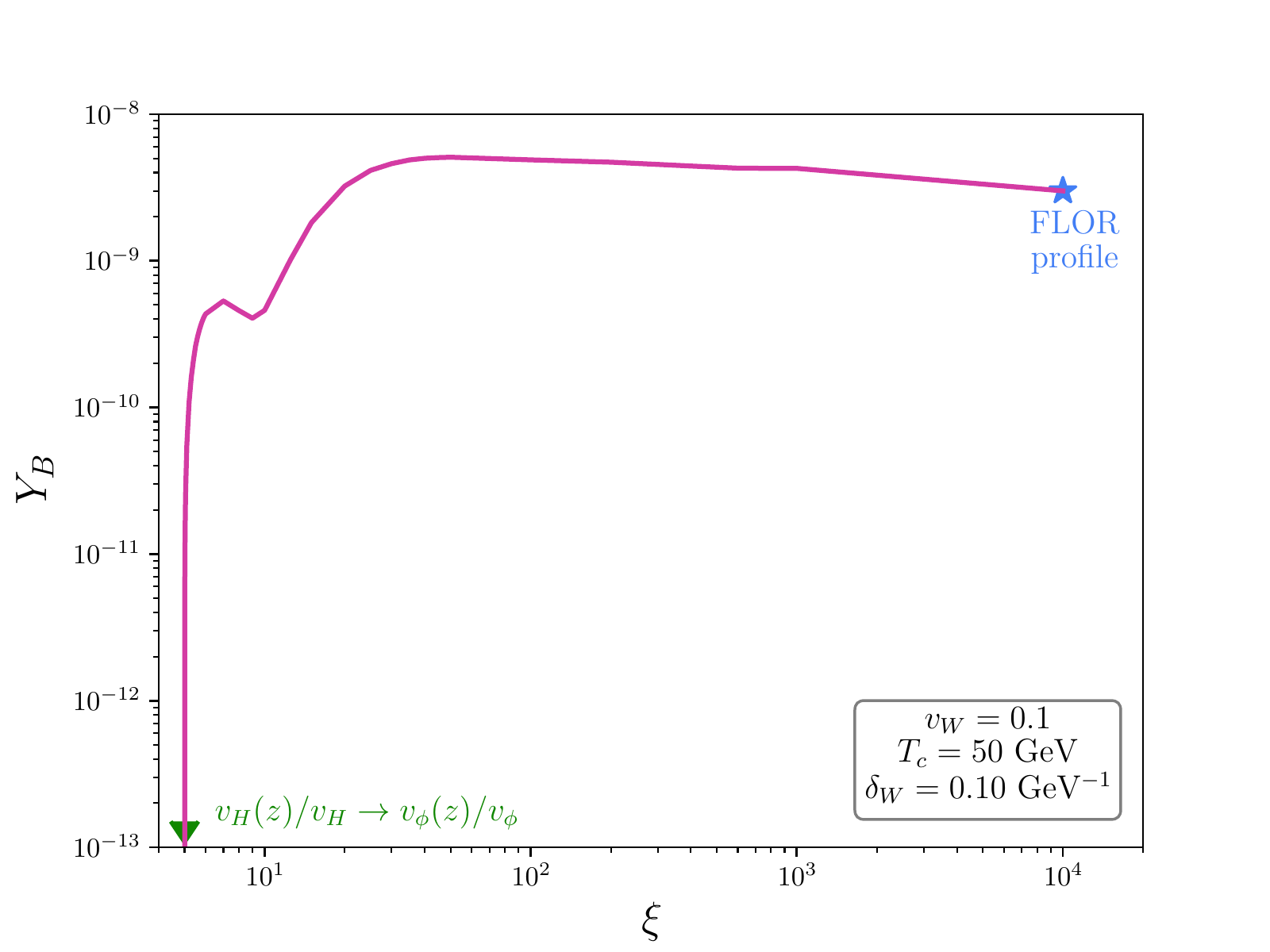}
    \caption{Generated $Y_B$ as a function of $\xi$ as defined in Eq.~(\ref{Eq:Profile_Tanh}) for a benchmark point. The blue star corresponds to the asymmetry generated using the FLOR profile, while the green triangle points towards the limiting case in which both vev profiles have the same shape where no asymmetry is generated. The masses of the heavy neutrinos in this particular case are $M_1=80$
   ~GeV, $M_2=90$~GeV and $M_3=100$~GeV while $m_D$ has been fixed as discussed at the end of Section~\ref{Sec:Lagrangian}. The heavy-active mixing has been set to $Tr[\theta\theta^\dagger]=0.007$~\cite{Fernandez-Martinez:2016lgt}.}
    \label{Fig:BAU_Steep_Tanh}
\end{figure}

Finally, we show in Fig.~\ref{Fig:BAU_Steep_Tanh} the final BAU for a benchmark point with $M_1=80$~GeV, $M_2=90$~GeV, and $M_3=100$~GeV, using the profiles from Eq.~(\ref{Eq:Profile_Tanh}) as a function of the $\xi$ parameter controlling the steepness of the Higgs vev within the wall. As can be noted, when using the kink profiles a slightly larger baryon asymmetry is generated with respect to the one obtained using the FLOR profile (blue star) when $\xi$ is large. However, as $\xi$ becomes smaller and the $z$-dependence of both profiles is more similar, the generated BAU starts shrinking until it becomes zero when $v_H(z)/v_H\rightarrow v_{\phi}(z)/v_{\phi}$, as expected.
\section{Conclusions}\label{Sec:Conclusions}
The origin of the matter-antimatter asymmetry in the Universe is still an open question in particle physics and cosmology. Several interesting possibilities have been proposed and developed in the literature to explain this unbalance. Among them, leptogenesis and electroweak baryogenesis are particularly compelling. 

Leptogenesis models have the appeal of connecting the generation of the baryon asymmetry with the mechanisms explaining neutrino masses, thus linking two experimental indications of new physics.
However, there are only few examples where these scenarios can be probed at present or near future facilities, rendering the mechanism essentially untestable in most cases.
Conversely, electroweak baryogenesis scenarios aim to explain the asymmetry through physics around the electroweak scale, which possibly relate to the Higgs hierarchy problem, making them much more testable, to the extent that measurements of electric dipole moments rule out many options. 
To avoid such constraints, it is typically necessary to include a dark sector with new sources of CP violation which generate the observed BAU while evading the tight EDM bounds. 

In this work we have studied the possibility that the mechanism responsible for neutrino masses also helps in the production of the BAU within the context of electroweak baryogenesis. 
Indeed, low scale realizations of the Seesaw mechanism such as the inverse or linear realizations, not only naturally explain the origin of tiny neutrino masses through an approximate lepton number symmetry, but are also testable since they allow for new heavy neutrinos at the electroweak scale with a sizeable mixing with their active partners. 
It is therefore tantalizing to explore the role of these new states and sources of CP violation at the electroweak scale in electroweak baryogenesis, particularly because the neutrino sector naturally avoids the problematic EDM constraints. 

This idea was first studied in Ref.~\cite{Hernandez:1996bu}, which was the starting point of our analysis that we expanded in several aspects, such as the impact of the vev profiles and the inclusion of flavour-dependent wash-out effects. 
In particular, we notice that the vev profiles studied in Ref.~\cite{Hernandez:1996bu} are rather conservative and tend to lead to a smaller BAU since the heavy-active neutrino mixing in the bubble wall is strictly smaller than its value in the broken phase, for which stringent constraints from flavour and electroweak precision observables apply. Indeed, the CP-invariant is proportional to the sixth power of this mixing and thus this choice critically impacts the final BAU asymmetry that may be obtained. In fact, upon solving the same set of diffusion equations described in Ref.~\cite{Hernandez:1996bu} where only the sphaleron process is included, we find that, even assuming the most suitable choices of the vev profiles, the observed BAU cannot be explained in this ``vanilla'' scenario due to the tight bounds on heavy-active neutrino mixing~\cite{Fernandez-Martinez:2016lgt}. However, if the singlet neutrinos were lighter than the $Z$ boson and decayed invisibly to a dark sector, some of these bounds would be sufficiently relaxed to allow the generation of the observed BAU in some small regions of the parameter space.

Next, we studied in detail the effect of including the interactions between the right-handed and SM neutrinos mediated by their Yukawa couplings in the final BAU. We find a very significant enhancement with respect to the vanilla scenario. 
Indeed, the GIM cancellation that takes place when adding the asymmetries from the different neutrino flavours is prevented by the different wash-out rates that each of them would have from their Yukawa interactions. Moreover, the asymmetries induced in the right-handed neutrino sector diffuse much farther from the bubble wall, given their weaker interactions, and can be transferred to the SM neutrinos and to baryons via the Yukawa and sphaleron processes, respectively. Thus, we find it is indeed possible to explain the observed BAU in agreement with current bounds on neutrino mixing when these effects are considered.

In this framework, the explanation of the observed BAU does require the extra neutrinos predicted by the low-scale Seesaw realizations to have masses around $100$~GeV and sizable mixing with the active neutrinos. This mechanism is thus potentially testable with collider searches 
and we leave a detailed exploration of the full parameter space as well as its detection prospects for future investigation. Another interesting open avenue of investigation is a detailed study of the scalar potential and the parameters characterizing the phase transition so as to ensure that suitable vev profiles are achievable. 

To summarize, we have studied two scenarios where the baryon asymmetry is generated from the CP violation stemming from the neutrino Yukawa couplings in a low-scale Seesaw mechanism. In the simplest case, neglecting the flavour-dependent Yukawa rates, we find it is not possible to explain the observed BAU unless present bounds on heavy-active neutrino mixing can be avoided. For instance, if the singlet neutrinos decay invisibly and are lighter than the $Z$ boson the constraints are sufficiently relaxed to achieve the observed BAU in a small window of the parameter space. More interestingly, when the flavour-dependent wash-out rates are included, the observed BAU can be successfully explained within present constraints. In any event, the required mixing is always large and these scenarios could be testable by future collider searches.

\section*{Acknowledgments}

The authors warmly thank Pilar Hern\'andez and Nuria Rius for very illuminating discussions and Jose Miguel No also for collaboration during the early stages of this work. EFM, TO and SRA  acknowledge the support of the Spanish Agencia Estatal de Investigacion and the EU ``Fondo Europeo de Desarrollo Regional'' (FEDER) through the projects PID2019-108892RB-I00/AEI/10.13039/501100011033 and FPA2016-78645-P as well as the``IFT Centro de Excelencia Severo Ochoa SEV-2016-0597''. JLP acknowledge the support from Generalitat Valenciana through the ``plan GenT'' program (CIDEGENT/2018/019) and from the Spanish MINECO under Grant FPA2017-85985-P.

\bibliographystyle{JHEP}
\bibliography{Baryogenesis_bib}

\end{document}